\documentclass[12 pt]{article}%%%%%%%%%%%%%%%%%%%%%%%%%%%%%%%%%%%%%%%%%%%%%%%%%%%%%%%%%%%%%%%%%%%%%%%%%%%%%%%%%%%%%%%%%%%%%%%%%%%%%%%%%%%%%%%%%%%%%%%%%%%%%%%%%%%%%%%%%%%%%%%%%%%%%%%%%%%%%%%%%%%%%%%%%%%%%%%%%%%%%%%%%%%%%%%%%%%%%%%%%%%%%%%%%%%%%%%%%%%%%%%%%%%%%%%%%%%%%%%%%%%%%%%%%%%%
\pdfoutput=1
\usepackage{fullpage}
\usepackage{amsmath}
\usepackage{authblk}
\usepackage{graphicx}
\usepackage{amsfonts}
\usepackage{amssymb}
\usepackage{natbib}
\usepackage[toc,page]{appendix}
\usepackage{sectsty}
\sectionfont{\centering}
\begin{document}

\title{\vspace{-2.0cm}Dynamics of Summer Monsoon Current around Sri Lanka}
	
	%\author {Subham Rath \footnote{%
		%Centre for Atmospheric and Oceanic Sciences, Indian Institute of Science,
		%Bangalore, India}, P. N. Vinaychandran\footnotemark[1], J. P. McCreary%
	%\footnote{%
		%International Pacific Research Center, University of Hawaii, Honolulu, Hawaii%
	%} and A. Behara\footnotemark[1]}
	%\Date
%	\author[add1]{Subham Rath}
%\author[add1]{P. N. Vinaychandran}
%%\ead{}
%\author[add2]{J. P. McCreary}
%%\ead{email3@domain.ca}
%\author[add1]{A. Behara}
%
%\address[add1]{Centre for Atmospheric and Oceanic Sciences, Indian Institute of Science, Bangalore, India}
%\address[add2]{International Pacific Research Center, University of Hawaii, Honolulu, Hawaii}
%\address[add3]{Department of Interests}

\author[1]{\small Subham Rath}
\author[1]{\small P. N. Vinaychandran}
%\author[2]{\small J. P. McCreary}
\author[1]{\small A. Behara}
\author[1]{\small C. P. Neema}

\affil[1]{\footnotesize Centre for Atmospheric and Oceanic Sciences, Indian Institute of Science, Bangalore, India}
%\affil[2]{\footnotesize International Pacific Research Center, University of Hawaii, Honolulu, Hawaii}
%\affil[3]{\footnotesize Department of Astrophysical Sciences, Princeton University, 4 Ivy Lane, Princeton, NJ 08544-1001 USA}
%\affil[4]{\footnotesize Department of Astronomy, Yale University, New Haven, CT 06520, USA}
%\affil[5]{\footnotesize Einstein and Spitzer Fellow}	
	\maketitle
	%\begin{abstract}%
		%EndExpansion
		%\setoutlinelevel2
		
		\section*{Abstract}
From June--September, the summer monsoon current (SMC) flows eastward south
of Sri Lanka and bends northeastward to form a swift jet that enters the Bay
of Bengal (BoB). As such, it is a crucial part of the water exchange
between the Arabian Sea (AS) and BoB. The processes that determine the
evolution, intensification and meandering of the SMC are only partly
understood. They involve both local and remote forcing by the wind, as well
as interactions with westward-propagating Rossby waves and eddies. In this study, we investigate these processes using an Indian-Ocean general
circulation model (MOM4p1) that is capable of simulating the SMC\
realistcally. Because eddies and meanders are smoothed out in the climatology, our analyses focus on a single year of 2009, a period when a strong anticyclonic bend in the SMC was observed.\par
An eddy-kinetic-energy budget analysis shows the region to be a zone
of significant eddy activity, where both barotropic and baroclinic
instabilities are active. Based on the analysis, we classify the evolution of
SMC into stages of onset, intensification, anticyclonic bend, anticyclonic
vortex formation, meandering and termination. In addition, analysis of
eddy-potential-vorticity flux and eddy-enstrophy decay reveal when, where,
and how the eddies tend to drive the mean flow. Rossby waves and
westward-propagating eddies arriving from the east energize the SMC in June
and accelerate the mean flow through an up-gradient eddy-potential-vorticity
flux. At the same time, local winds also strengthen the flow, by increasing its
mean, near-surface, kinetic energy and raising isopycnals, the latter
building up available potential energy (APE). The baroclinic instability
that takes place in late--July and early--August releases APE, thereby
generating the SMC meanders.

		%\end{abstract}%

%\setoutlinelevel3
\clearpage
	
	\section{Introduction\label{introduction}}

The Indian--Ocean (IO) monsoon currents flow south of Sri Lanka, flowing eastward during the summer (named the Summer Monsoon Current;
SMC) and westward during the winter (named the Winter Monsoon Current; WMC) [%
\citealp{schott2001monsoon}; \citealp{shankar2002monsoon}].The SMC plays a
particularly important role in the transfer of more saline water from
the Arabian Sea (AS) to the Bay of Bengal (BoB)( \citealt{murty1992water}, %
\citealt{vinayachandran1999intrusion}, \citealt{jensen2001arabian}).
\subsection{Background}
\begin{figure}[h]
\centering
\includegraphics[width=1\linewidth]{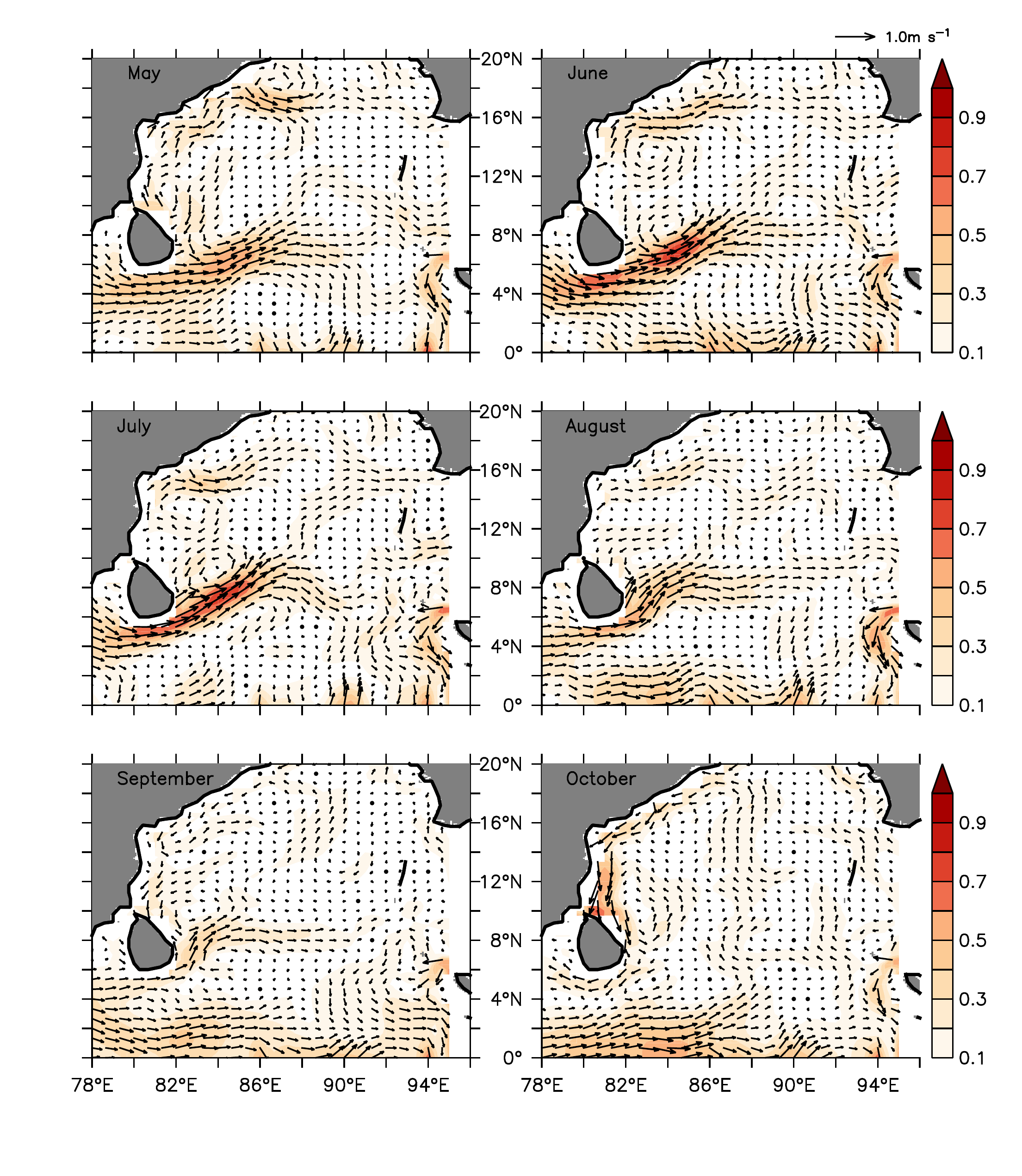}
\caption{Monthly maps of climatological (2000--2009) currents from OSCAR averaged from May--October. The scale vector (1 m/s) is shown above panel.}
\label{fig:oscar_climatology1}
\end{figure}

	%\subsubsection{Observations}
	%\paragraph{OSCAR monthly climatology}
	Figure \ref{fig:oscar_climatology1} illustrates the SMC, providing bimonthly
plots of the circulation climatology from the Real-time Ocean Surface Current Analysis (OSCAR). The SMC intensifies during May, attains speeds higher than the currents in the other parts of the IO, and makes a northeastward
turn into the BoB. There are cyclonic and anticyclonic circulations on the
west and east of the strong current, respectively. This pattern remain during July and August, and diminishes in September. It appears every year,
albeit with some variation in location and timing. \par 
By compiling data obtained from satellite-tracked drifting buoys, %
\citet{molinari1990surface} first reported eastward flow during the summer
near 10$^{\circ }$N in the western and central Indian Ocean, naming it the SMC. 

  \citet{hastenrath1991monsoonal} described the SMC as being primarily wind
driven. The first current-meter measurements in the region revealed the SMC to
be shallow \citep{schott1994currents}. Using satellite-tracked drifting
buoys, \citet{shenoi1999near} reported that the SMC branched around 87$%
^{\circ }$E, with one branch entering the BoB and the other continuing eastward. Expendable bathythermograph (XBT) observations collected between
Sri Lanka and the Malaca Strait, together with altimeter data and Oceanic General Circulation Model (OGCM) results, showed the current bends around Sri Lanka and flows into the BoB %
\citep{vinayachandran1999intrusion}.
	
%\paragraph{Eddy activity}
\par 	
Significant eddy activity was also reported in the region during the
summer monsoon, based on both drifting-buoy data \citep{shenoi1999near} and 17
years of altimetry data \citep{chen2012features}.
Large barotropic and baroclinic energy conversions are associated with regions of high eddy kinetic energy, suggesting the importance of instabilities in eddy generation %
\citep{cheng2013intraseasonal}. \citet{cheng2013intraseasonal} also
commented on the intensification of SMC, suggesting that its probable cause
was the conversion of eddy-to-mean kinetic energy during June.

	%\paragraph{Recent observations}
	\par 
	Recently, an international team carried out extensive measurements
	to investigate monsoon-current dynamics south of Sri Lanka %
	\citep{lee2016collaborative}. Observations from satellite remote sensing,
	gliders, moorings, drifters, and echo sounders were combined to present a
	more comprehensive view of the circulation. After crossing the southern tip
	of Sri Lanka, the SMC flows around the perimeter of a distinct region of
	shallow thermocline, the Sri Lanka dome (SLD; %
	\citealp{vinayachandran1998monsoon}). \citet{lee2016collaborative} also noted strong northward surface  geostrophic flow along 8$^{\circ }$N between 83$^{\circ }$E and 84$^{\circ }$E.
	
	%\subsubsection{Models}
	\par 
	A number of modelling studies have investigated SMC dynamics. In companion
	papers, \citet{shankar1996dynamics} and \citet{mccreary1996dynamics} found a
	southward flow of East Indian Coastal Current (EICC), east of Sri Lanka
	during the summer, surprising since the current flows \textit{against} the
	southerly alongshore winds. They determined that the southward flow occurs
	due to local forcing by a region of strong Ekman suction just east of Sri
	Lanka.
	%\subparagraph{Rossby waves and SL dome}
	\par 
	McCreary et al.\ (1993) and \citet{vinayachandran1998monsoon} showed that
	the northward turning is influenced (enhanced) by the arrival of a Rossby
	wave associated with northward flow from the eastern boundary. %
	\citet{vinayachandran1998monsoon} noted that both Ekman pumping and
	Rossby-wave arrival determined the life cycle of SLD. A large anticyclonic
	vortex was found to exist south of Sri Lanka along with an anticyclonic eddy
	east of SLD; the anticyclonic vortex was formed as a result of barotropic
	instability. Further, it was found that the arrival of a second Rossby-wave
	signal (associated with southwestward flow) terminated the northward turning.

%\paragraph{Impact of intraseasonal variability and eddies}
 \par  
Recently, a study by \citet{cheng2017intraseasonal} on the intraseasonal- to-
semiannual variability of sea surface height (SSH) in the southern BoB revealed strong 30-60 day variability east of Sri Lanka.This variability was primarily caused by nonlinear Rossby waves from the east. Further,
they showed that the variability at periods greater than 60 days nonlinearly
transferred energy to the 30--60-day band.

\subsection{Present research}
	%\subsubsection{Goals}
	In this paper, we continue the effort to understand SMC dynamics. In this study, we seek to understand the processes that are involved in the interaction between the SMC and
eddies, and in causing unusually high SMC speed, the meandering of the current, and the ultimate termination of SMC in the
southeastern BoB. The relative importance of winds and westward propagating
eddies in modulating the mean flow in its various stages is also
investigated.\par 
Toward these goals, we analyze a solution to an OGCM which represents the region's circulation realistically. Eddy-kinetic-energy budget and Transformed Eulerian Mean
framework are used to identify regions of eddy-mean flow interaction.
To quantify the impacts of local winds, mean wind power
integrated over the region of interest is compared with volume
integrated mean kinetic energy of SMC above the mixed layer depth. The two
quantities are related to each other by the mean kinetic energy equation
(section 3). The geostrophic wind power, which is utilised in raising the
isopycnals, is also integrated over the study region and compared to the integrated available potential energy (APE).
	
	The mean current circulation in summer (June--September) from during 2004--2009 (Fig.\ \ref%
{fig:inter_annual_smc1}) shows the expected north eastward bending of SMC as
discussed in \citet{vinayachandran1998monsoon}. In addition, the SMC is observed
to meander and feed into the southeastern BoB in each of the six years. The climatological
map fails to show the meandering of the current. Therefore, we choose a single year of, 2009, for our
analyses. This choice is primarily motivated by the study of Vinayachandran
et al.\ (2013), which showed a sharp anticyclonic bend of the current during
that year. 
\subsubsection{Key results}

Our study divides the evolution of SMC during June--September into six stages-- onset, intensification, anticyclonic bending, anticyclonic vortices, meandering and termination. Energy exchanges between background flow and the eddies during these stages are examined.
Among these six stages, of particular interest are the stages of intensification, anticyclonic bending, and meandering of SMC. Our work, with the help of energy budget and Transformed Eulerian mean (TEM) approach shows that the effect of eddies in intensifying the SMC is localized to a small region. We found that the strengthening of local winds was primarily responsible for intensification of SMC. The anticyclonic turning of the SMC in early July occured in response to the local winds as well. Further, the meandering of the SMC was as a result of baroclinic instability (conversion of APE to eddy kinetic energy) which was caused due to fluctuating wind power.  \par 
 
 %\subsubsection{Organization}
	The paper is organized as follows. In Section \ref{modelandsolution}, we provide model overview and compare model output with observations.  Section 3 is based on eddy kinetic-energy-budget analysis. In Section 4, we present the evolution
of SMC and discusses its five stages with the help of energetics and Transformed Eulerian mean (TEM) approach. Also, investigate the role of the local winds in
these stages. In Section 5, a summary of our findings is provided.
	
	\section{The model \label{modelandsolution}}
	
	%\subsection{Model overview}
	
The OGCM used in our study is the Modular Ocean Model (MOM), a numerical
ocean model based on hydrostatic primitive equations. The model
configuration, forcings as well as boundary conditions used
in this study are the same as in \citet{das2015formation}. So, only a brief
overview is given here.
	\par 
	The model domain is the tropical IO, extending from 30$^{\circ }$S to
30$^{\circ }$N and from 30--120$^{\circ }$E. Its horizontal resolution is
0.25$^{\circ }$ in both latitude and longitude. It has
variable resolutions in the vertical with 58 levels, 44 of which
are in the top 200 m, and 1-m resolution is used in the top 18 m. Bottom
topography is based on modified ETOPO5 bathymetry %
\citep{sindhu2007improved}. Continental boundaries within the domain are
treated as solid walls, and sponge layers are applied long the southern model boundary and along portions of the eastern
boundary \citep{kurian2007mechanisms}. Vertical mixing is the
K-profile parametrization (KPP) scheme \citep{large1994oceanic}, and
horizontal mixing scheme follows \citet{chassignet2001viscosity}. Vertical
diffusivity and viscosity are based on dissipation mechanisms associated
with internal wave breaking \citep{simmons2004tidally} and bottom drag
experienced by barotropic tides \citep{lee2006rectification}, respectively. The model
equation of state is the same as that in \citet{jackett2006algorithms}.
The model is spun up for 10 years using climatological forcing, which is
then switched to daily varying interannual forcing (see %
\citet{vinayachandran2012mechanisms} for details). 
\par
\begin{figure}[h]
\centering
\includegraphics[width=1\linewidth]{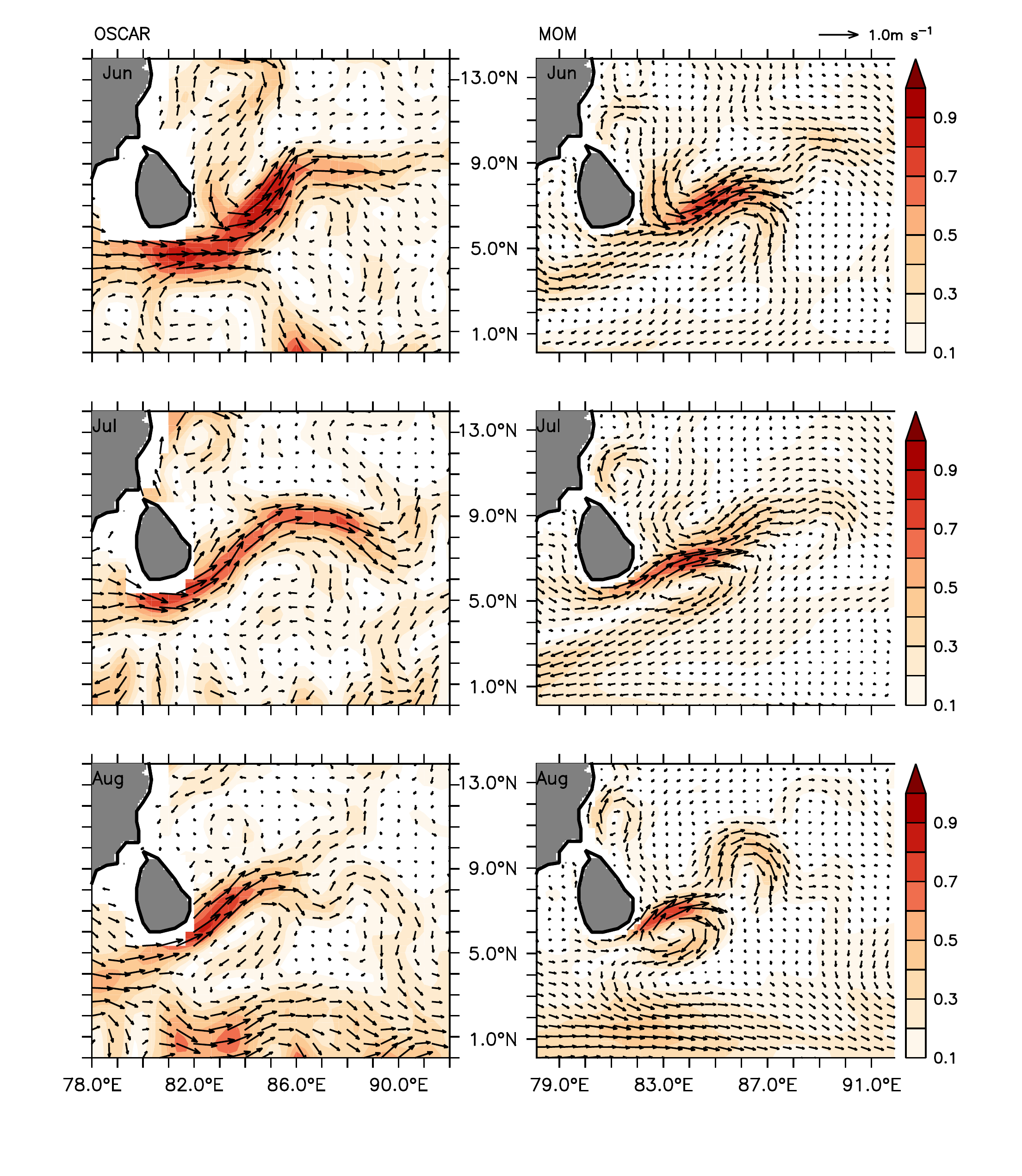}
\caption{Comparison of monthly maps of currents of southwest monsoon, 2009 from OSCAR (left panels), MOM (right panels). Current vectors overlaid over speed (shading, cm/s). Vectors are of same length in all panels and the scale vector (1m/s) is shown above panel.}
\label{fig:oscar_mom_com}
\end{figure}

 Figure \ref{fig:oscar_mom_com} compares observed and modelled, surface
currents during June, July, and August of 2009. Large-scale features of the
circulation are well simulated, except that the model underestimates the current
speed. Small scale features, such as
anticyclonic eddies during July and August, were more prominent in the
observations. For example, during July and August the OSCAR product show a continuous
meandering of SMC, whereas an anticyclonic vortex is found
together with the SMC as it entered the BoB in the model. During August, two
vortices were present to the right of the SMC in the model. The northern vortex in the model did not exist in the observation.

		%\setoutlinelevel3
	
	\section{Eddy activity\label{eddyactivity}}
	
\subsection{Mean and eddy kinetic energy\label{MKE&EKE}}
	
To investigate the eddy-mean flow interactions that impacts the SMC, we
divide all the field variables of interest into a background (mean) and eddy component. In our study, we have taken 120 day low-pass filtered state as the mean component and 120 day high-pass filtered state as the eddy component (see Appendix A). The kinetic energy is also split into mean kinetic energy (MKE) and eddy kinetic energy (EKE).  Following \citep{masina1999analysis} notation, MKE is defined by %
\begin{equation}
K=\rho _{0}\frac{\bar{u_{i}^2}}{2}\equiv \rho _{0}\left( \frac{ \bar{u^2} +%
\bar{v^2}}{2}\right) ,  \label{eqn:Kdef}
\end{equation}
and determined by the prognostic equation 
\begin{equation}
\begin{array}{c}
\underbrace{\dfrac{\partial }{\partial t}K}_{A}+\underbrace{\rho _{0}u_j 
\dfrac{\partial }{ \partial x_{j}}\dfrac{\overline{u_{i}^{2}}}{2}}_{B}=%
\underbrace{-\dfrac{\partial }{\partial x_{i}}\overline{ Pu_{i}}}_{C} -%
\underbrace{\rho _{0}\overline{u_{i}}\dfrac{\partial \overline{u_{i}^{\prime
}u_{j}^{\prime }}}{\partial x_{j}}}_{D} \underbrace{-g\bar{\rho }\bar{w}}%
_{E} \underbrace{+\overline{u_{i}}\cdot\overline{\mathcal{D}}}_{F}+%
\underbrace{\overline{u_{i}}\cdot\overline{\mathcal{F}}}_{G}.%
\end{array}
\label{eqn:Keqn}
\end{equation}

\begin{itemize}
  %\hline 
  \item[] A : MKE tendency
  \item[] B : Advection of MKE 
  \item[] C : Pressure divergence flux 
  \item[] D : Eddy momentum flux convergence
  \item[] E : Conversion mean potential energy (MPE) to MKE
  \item[] F : Dissipation
  \item[] G : MKE change due to external forcing
 \end{itemize}

\par Similarly, EKE is defined by%
	\begin{equation}
	K^{\prime}=\rho _{0}\frac{u_{i}^{\prime 2}}{2}\equiv \rho _{0}\left( \frac{%
		u^{\prime 2}+v^{\prime 2}}{2}\right) ,  \label{eqn:Kedef}
	\end{equation}%
	and determined by the prognostic equation%
	\begin{equation}
	\begin{array}{c}
	\underbrace{\dfrac{\partial }{\partial t}K^{\prime}}_{A^{\prime }}+\underbrace{\rho
		_{0}\left( \overline{u_{j}}+u_{j}^{\prime }\right) \dfrac{\partial }{%
			\partial x_{j}}\dfrac{u_{i}^{\prime 2}}{2}}_{B^{\prime }}=\underbrace{-\dfrac{\partial 
		}{\partial x_{i}}\left( P^{\prime }u_{i}^{\prime }\right) }_{C^{\prime }}
	\underbrace{-\rho _{0}u_{i}^{\prime }u_{j}^{\prime }\dfrac{\partial 
			\overline{u_{i}}}{\partial x_{j}}}_{D^{\prime }}-\underbrace{g\rho ^{\prime }w^{\prime }}_{E^{\prime }}+\underbrace{u_{i}^{\prime}\cdot\mathcal{D}^{\prime}}_{F^{\prime }} +\underbrace{u_{i}^{\prime}\cdot\mathcal{F}^{\prime}}_{G^{\prime }}.%
	\end{array}
	\label{eqn:Keeqn}
	\end{equation}%
	%\clearpage
	\begin{itemize}
   %\hline 
  \item[] $A^{\prime}$ : EKE tendency
  \item[] $B^{\prime}$ : Advection of EKE
   \item[] $C^{\prime}$ : Eddy pressure divergence flux
   \item[] $D^{\prime}$ : Deformation work due to mean flow
   \item[] $E^{\prime}$ : Baroclinic energy conversion
   \item[] $F^{\prime}$ : Eddy disipation
   \item[] $G^{\prime}$ : EKE change due to external forcing
   \end{itemize}

%	The terms on the left-hand side of (\ref{eqn:Keeqn}) represent the local (A)
%	and advective (B) rates of change of $K_{e}$. The terms on the right-hand
%	side are rates of change of EKE due to: pressure work (C); viscous-stress
%	transport and viscous dissipation (D); deformation work due to mean flow
%	(E); eddy dissipation term (F); and buoyancy production (G). In large-scale
%	flows, terms (D) are insignificant, and so we neglect them in our analyses.
	
%	Eddy pressure divergence flux term appearing in (4) tends to redistribute
%	energy. Positive value of $-\boldsymbol{\nabla }\cdot (P^{\prime }%
%	\boldsymbol{u}^{\prime })$ in time mean sense, implies it to be a possible
%	source of EKE and tends to radiate energy from regions where eddy kinetic
%	energy generation is characterised by baroclinic or barotropic
%	instabilities. Time mean vertically integrated maps of eddy pressure
%	divergence flux (not shown here) implies horizontal divergence flux to be
%	the dominant term relative to the vertical divergence flux term. Alternate
%	sources and sinks of energy over the region possibly reflect the exchange
%	and confinement of energy radiated within the region of interest. Advection
%	terms are found to be of the same order and redistribute energy radiated
%	from instabilities within the region.
	
\subsubsection{EKE structure\label{EKEstructure}}
	
\begin{figure}[h]
\centering
\includegraphics[width=1\linewidth]{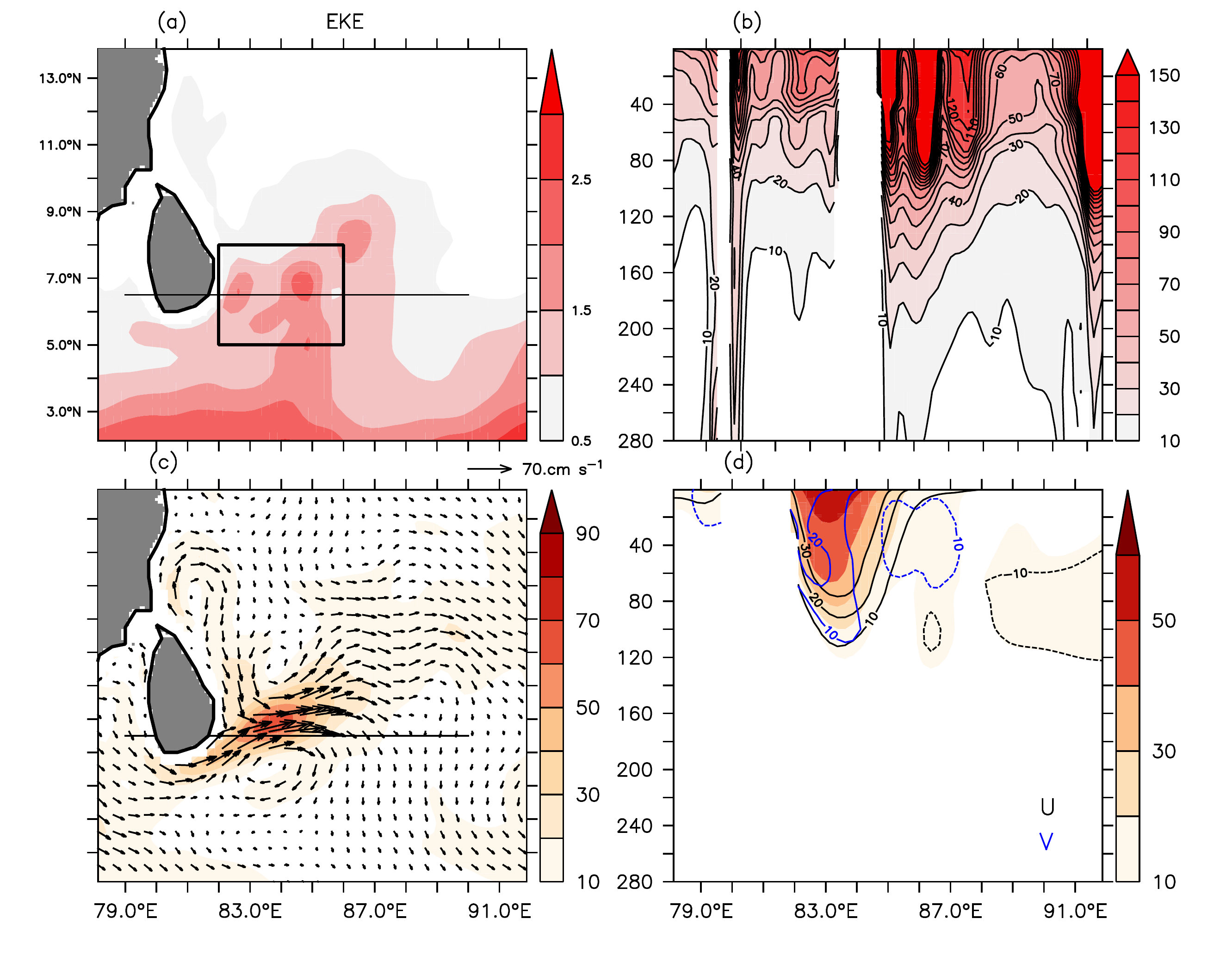}
\caption{(a) Time mean (June-September) vertically integrated EKE map (cm$^{3}$/s$^{2}$); (b) depth longitude section along 6.5$^{\circ }$N of time mean (Jun-Sep) EKE (10$^{-4}$ cm$^{2}$/s$^{2}$). Vectors are of length 60 cm/s. (c) Time mean (June-September) maps of surface averaged (30m) current vectors from the model overlaid over speed (shading cm/s); (d) time mean depth longitude sections (6.5$^{\circ }$N) of zonal velocity (red), meridional velocity (blue) contour spacing of 10cm/s overlaid over speed ofcurrent (cm/s) .}
\label{fig:paper_current_eke}
\end{figure}

	Figures \ref{fig:paper_current_eke}a and \ref{fig:paper_current_eke}b show
the time mean (June--September) vertically integrated (top 100 m) map of EKE
and depth-longitude section at 6.5$^{\circ }$N, respectively. The magnitude of EKE is consistent with the findings of \citet{chen2012features}. Maximum
EKE appeared between 83--85$^{\circ }$E in the form of
two distinct high-value cores extending to 120 m. Figures \ref%
{fig:paper_current_eke}c and \ref{fig:paper_current_eke}d show the surface-layer
averaged (top 30 m) time mean velocity maps and depth-longitude section of
velocities at 6.5$^{\circ }$N, respectively. High speeds of SMC existed between 83--85$^{\circ }$E and extended up to a depth of 120 m, consistent
with the shallow nature of SMC in the observation.
	
\subsubsection{Time series\label{MKE&EKEtimeseries}}

\begin{figure}[h]
\centering
\includegraphics[width=1\linewidth]{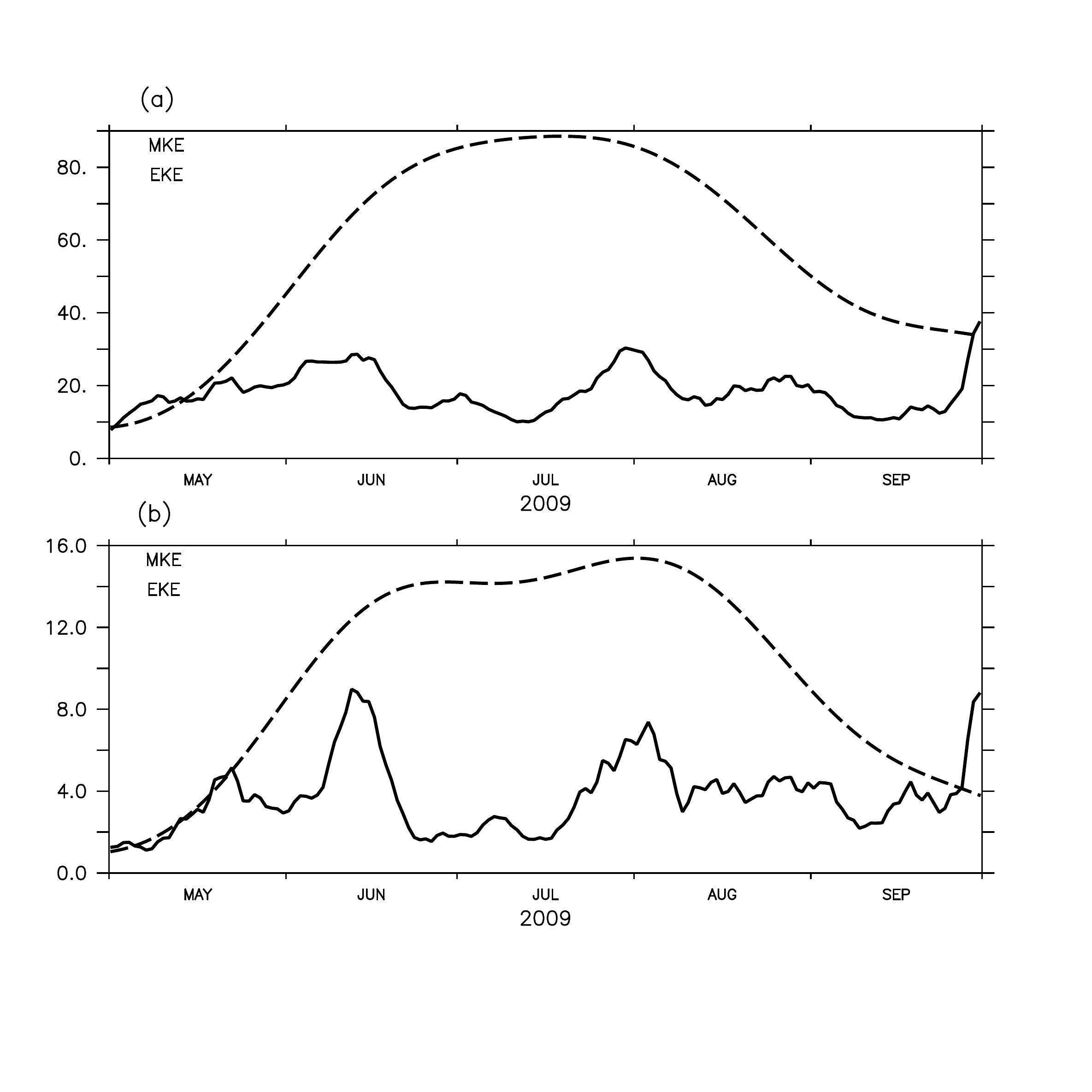}
\caption{(a) A volume integrated time MKE (dashed lines) and EKE (solid lines)
		between 82--86$^{\circ }$E, 5--8$^{\circ }$N, 0-100m. Units-- ($
		10^{1}{}^{6}cm^{2}/s^{2}$); (b) volume integrated time MKE and EKE between
		84--85$^{\circ }$E and 5--7$^{\circ }$N.}
\label{fig:paper_MKE_EKE_1}
\end{figure}

Figure \ref{fig:paper_MKE_EKE_1} shows the time series of volume-integrated MKE and EKE between 82--86$^{\circ
}$E, 5--8$^{\circ }$N in the top 100 m. The MKE is about four times larger in
magnitude than the EKE (Fig.\ \ref{fig:paper_MKE_EKE_1}a). This
difference shows eddies only had a secondary contribution in driving the
mean flow, which implies that the mean flow was largely wind driven when
integrated over a large domain. However, the effect of eddies in driving
the mean flow became more significant when integrated over a smaller domain
(84--85$^{\circ }$E and 5--7$^{\circ }$N) as the peak MKE was twice the peak
EKE now (Fig.\ \ref{fig:paper_MKE_EKE_1}b). Thus, fractional transfer from EKE to MKE is unlikely to bring in significant changes in SMC
speed throughout the domain, but the effects at smaller scales are prominent
as will be discussed later (Section\ 4.2.1).
	
		\subsection{Instabilities\label{instabilities}}
	
	Instabilities are associated with energy conversions. 
 Barotropic instability converts MKE to EKE, and is represented by term $D^{\prime}$ in (%
	\ref{eqn:Keeqn}), which when expanded has the form%
	\begin{equation}
	-\rho _{0}u_{i}^{\prime }u_{j}^{\prime }\frac{\partial \overline{u_{i}}}{%
		\partial x_{j}}=-\rho _{0}\left( u^{\prime }v^{\prime }\frac{\partial 
		\overline{u}}{\partial y}+v^{\prime }v^{\prime }\frac{\partial \overline{v}}{%
		\partial y}+u^{\prime }u^{\prime }\frac{\partial \overline{u}}{\partial x}%
	+u^{\prime }v^{\prime }\frac{\partial \overline{v}}{\partial x}\right) .
	\label{eqn:barotropicECTs}
	\end{equation}%
	Since the terms in (\ref{eqn:barotropicECTs}) are of the same magnitude  all are retained in our analyses. Baroclinic instability converts
APE into EKE. It is represented by the term $E^{\prime}$ ($%
	-g\rho^{\prime} w^{\prime }$) in (\ref{eqn:Keeqn}). For both instabilities, positive
	value of each terms implies energy conversion from the background state to eddies.
	
	\subsubsection{Horizontal and vertical structures}
	
\begin{figure}[h]
\centering
\includegraphics[width=\linewidth]{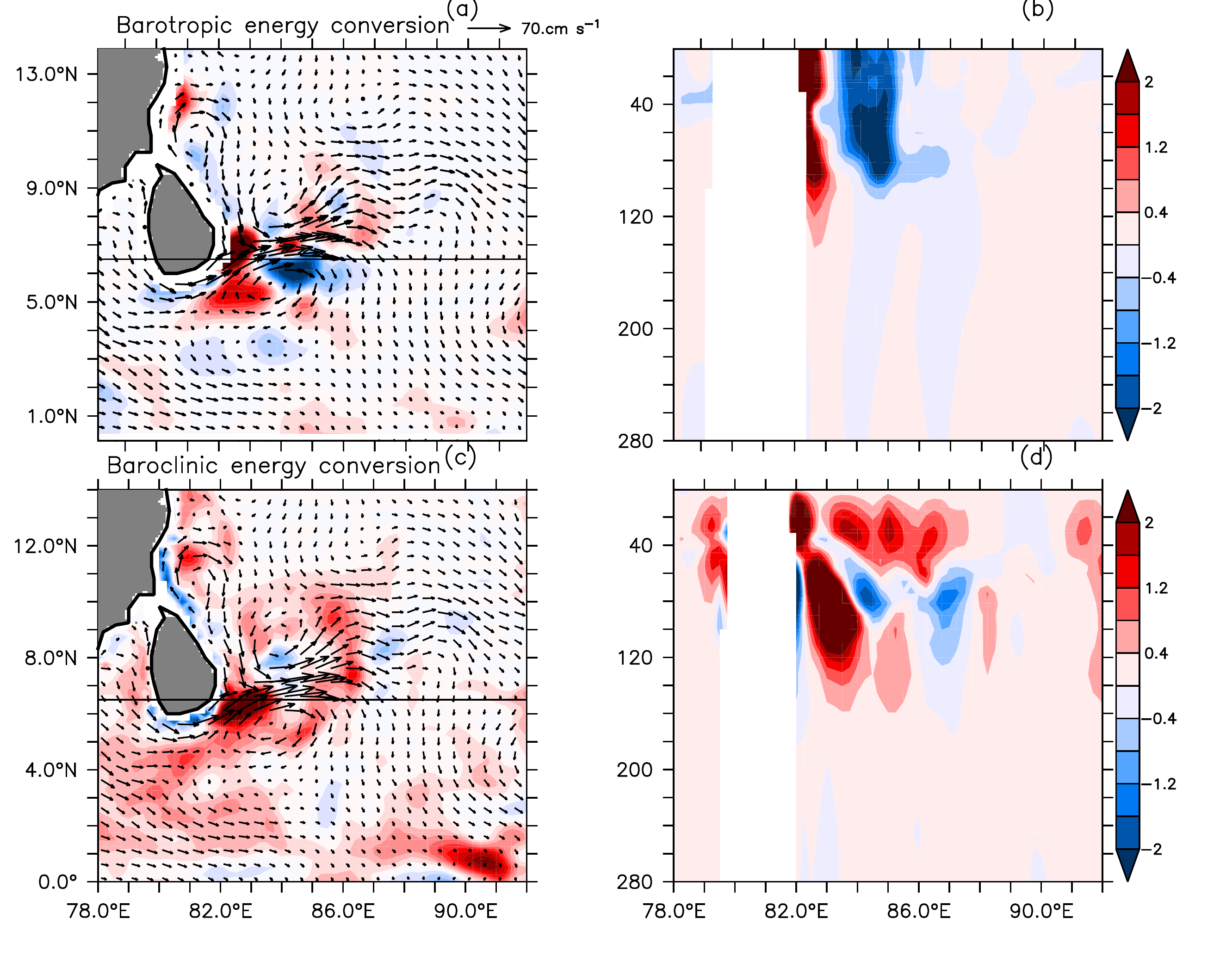}
\caption{Time mean (June--September) vertically integrated (0-100 m) map of (a) barotropic; (c) baroclinic energy conversion terms (cm$^{3}$/s$^{3}$). Time mean depth longitude sections ($10^{-4}$ cm$^{2}$/s$^{3}$) of (b) barotropic energy conversion, (d) baroclinic energy conversion at 6.5$^{\circ }$N. Time mean current vectors are averaged over 50 m are of same scale (0.6m/s) in both panels.}
\label{fig:paper_energy_conv}
\end{figure}

	Figures \ref{fig:paper_energy_conv}a and, \ref{fig:paper_energy_conv}c show time
mean vertically integrated (top 100 m) maps of barotropic and baroclinic energy
conversions, respectively. Figures \ref{fig:paper_energy_conv}b and \ref%
{fig:paper_energy_conv}d are the respective depth-longitude sections
of the energy conversions at 6.5$^{\circ }$N. Barotropic energy involving
both mean-eddy and eddy-mean energy conversions between (83$^{\circ }$E--85%
$^{\circ }$E) and (5$^{\circ }$N--7$^{\circ }$N). Eddies are found to
extract energy out of the mean flow west of 84$^{\circ }$E through
barotropic instability and transfer energy to the mean flow east of 84$%
^{\circ }$E. Significant baroclinic energy conversion was also found (Figs.\ %
\ref{fig:paper_energy_conv}c, \ref{fig:paper_energy_conv}d), indicating that
the SMC is characterized by mixed instabilities. Depth--longitude section
along 6.5$^{\circ }$N reveals barotropic energy conversion from mean to
eddies, extending to 100 m, while the baroclinic energy conversion was maximum between 60 and 80 m centered around 84$^{\circ }$E.

	\subsection{Intraseasonal Rossby waves and eddies}
	
\begin{figure}[h]
\centering
\includegraphics[width=1\linewidth]{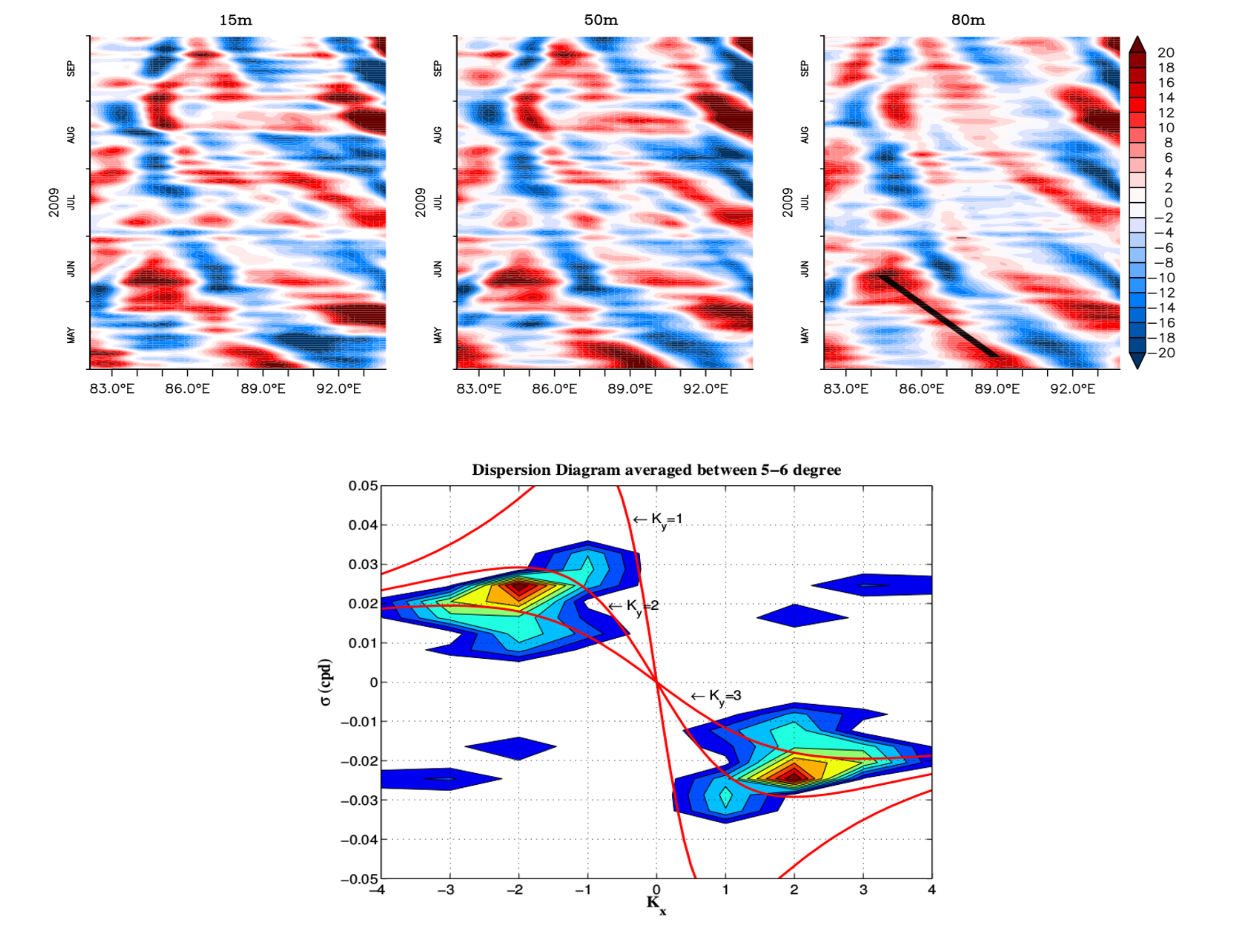}
\caption{Hovm{\H{o}}ller diagram (top) of high passed filtered meridional velocities averaged from 5--6$^{\circ }$N. Levels 15 m, 50 m, 80 m. Dispersion diagram (bottom) frequency ($\sigma $) versus wavenumber ($k$).}
\label{fig:paper_rossby}
\end{figure}

	 Figure.\ \ref{fig:paper_rossby} (top panel) is a Hovm{\"{o}}ller diagram of
high-pass-filtered meridional velocity. Westward-propagating signals with speeds
of about 15.5 cm/s started at 90$^{\circ }$E and propagated across the basin.
In particular, a signal associated with northward flow was at 90$^{\circ }$E
in May and reached 84$^{\circ }$E by mid-June. It was followed by a signal
with southward flow that arrived near 84$^{\circ }$E about mid-July. Note
that the propagation speed of the latter signal decelerated toward the west (the
band tilted more sharply), likely due to interaction with the SMC. 
\\
Dispersion diagram from high--pass-filtered meridional velocity component (Fig.\ \ref%
{fig:paper_rossby}) shows a probable Rossby wave structure with a 40--50 day
signal corresponding closely to baroclinic mode 2. The south and east of Sri Lanka
was characterized by the interactions of Rossby waves and other probable
eddies with the SMC and is consistent with the
findings of \citet{vinayachandran1998monsoon}. Rossby wave phase speed calculated from the dispersion
diagram was approximately 17 cm/s, consistent with the Hovm{\"{o}}ller diagram.

	\section{Evolution of SMC\label{evolutionofSMC}}
    
\begin{figure}[h]
\centering
\includegraphics[width=1\linewidth]{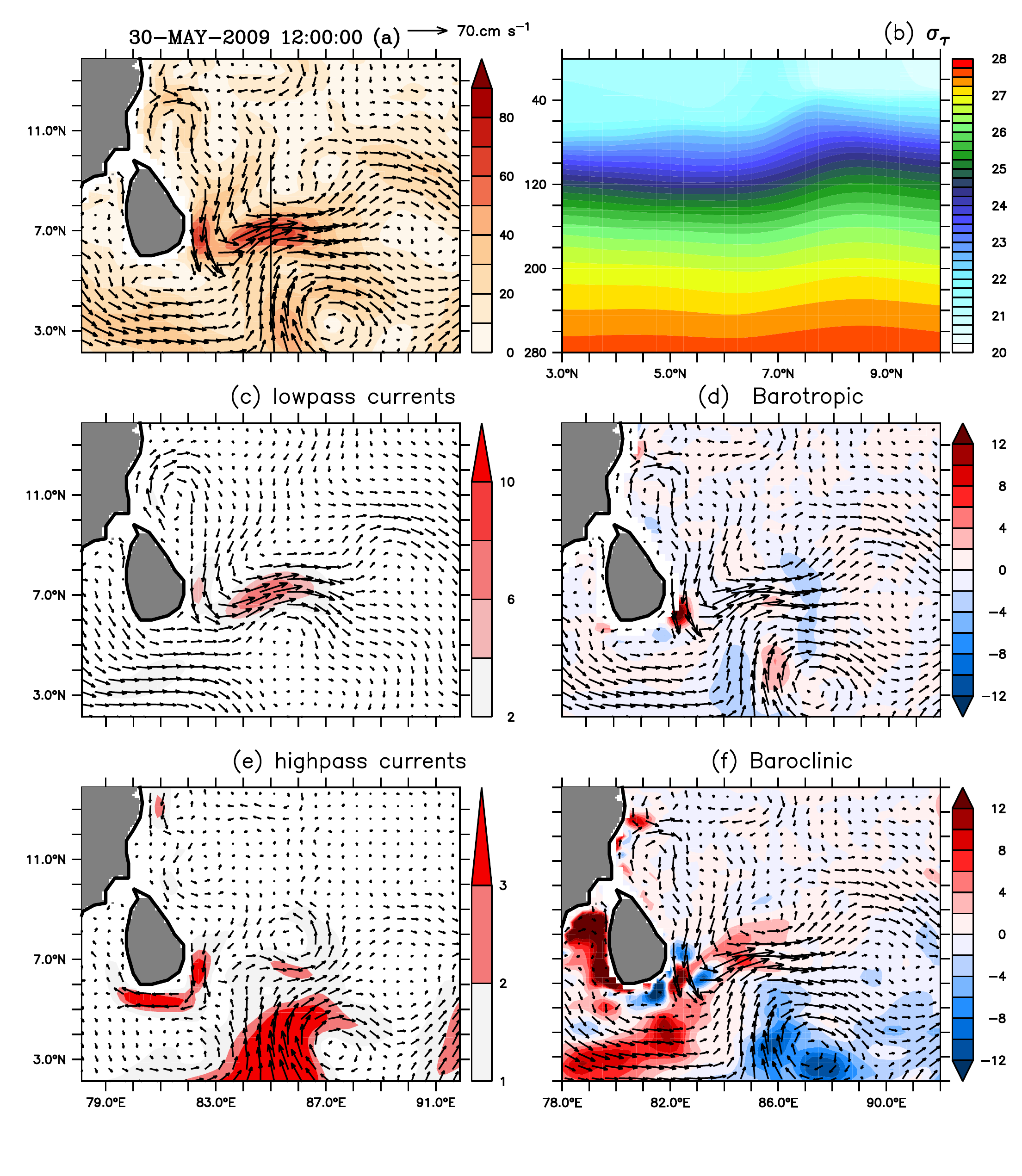}
\caption{(a) Current vectors overlaid over speed (shading, cm/s) averaged over top 30m; (b) $\sigma _{t}$ (kg/m$^{3}$); (c) 120 day lowpass filtered currents (top 30m depth averaged) as mean flow overlaid over top 100 m vertically integrated map of MKE (shading, cm$^{3}$/s$^{2}$); (d) Current vectors (top 30m depth averaged) overlaid over top 100 m vertically integrated map of barotropic energy conversion $-u_{i}^{\prime }u_{j}^{\prime }\partial \overline{u_{i}}/\partial y$ (shading, cm$^{3}$/s$^{3}$); (e) 120 day highpass filtered currents (cm/s) averaged over top 30 m, as eddies overlaid over top 100 m vertically integrated map of EKE (shading, $10^{6}$ cm$^{3}$/s  $^{2}$); (f) Current vectors (top 30m depth averaged) overlaid over top 100 m vertically integrated map of baroclinic energy conversion (shading, cm$ ^{3} $/s$^{3}$). All quatities are 5--day running averaged. Vectors are of same length in all panels and the scale vector is shown in above panel (a).}
\label{fig:crmultiplot_evo85n_150}
\end{figure}

\begin{figure}[h]
\centering
\includegraphics[width=1\linewidth]{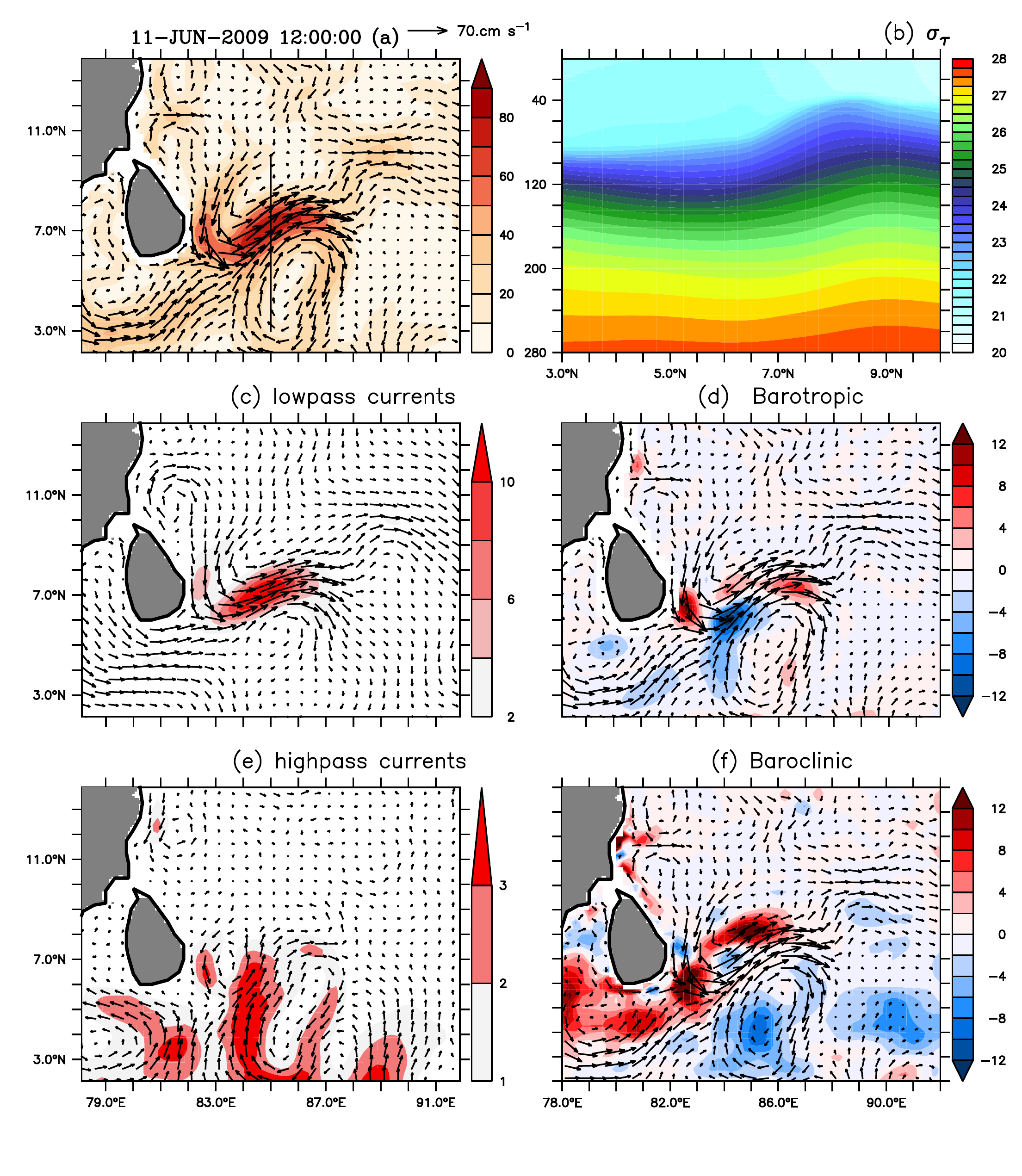}
\caption{(a) Current vectors overlaid over speed (shading, cm/s) averaged over top 30m; (b) $\sigma _{t}$ (kg/m$^{3}$); (c) 120 day lowpass filtered currents (top 30m depth averaged) as mean flow overlaid over top 100 m vertically integrated map of MKE (shading, cm$^{3}$/s$^{2}$); (d) Current vectors (top 30m depth averaged) overlaid over top 100 m vertically integrated map of barotropic energy conversion $-u_{i}^{\prime }u_{j}^{\prime }\partial \overline{u_{i}}/\partial y$ (shading, cm$^{3}$/s$^{3}$); (e) 120 day highpass filtered currents (cm/s) averaged over top 30 m, as eddies overlaid over top 100 m vertically integrated map of EKE (shading, $10^{6}$ cm$^{3}$/s $^{2}$); (f) Current vectors (top 30m depth averaged) overlaid over top 100 m vertically integrated map of baroclinic energy conversion (shading, cm$ ^{3} $/s$^{3}$). All quatities are 5--day running averaged. Vectors are of same length in all panels and the scale vector is shown in above panel (a).}
\label{fig:crmultiplot_evo85n_162}
\end{figure}

\begin{figure}[h]
\centering
\includegraphics[width=1\linewidth]{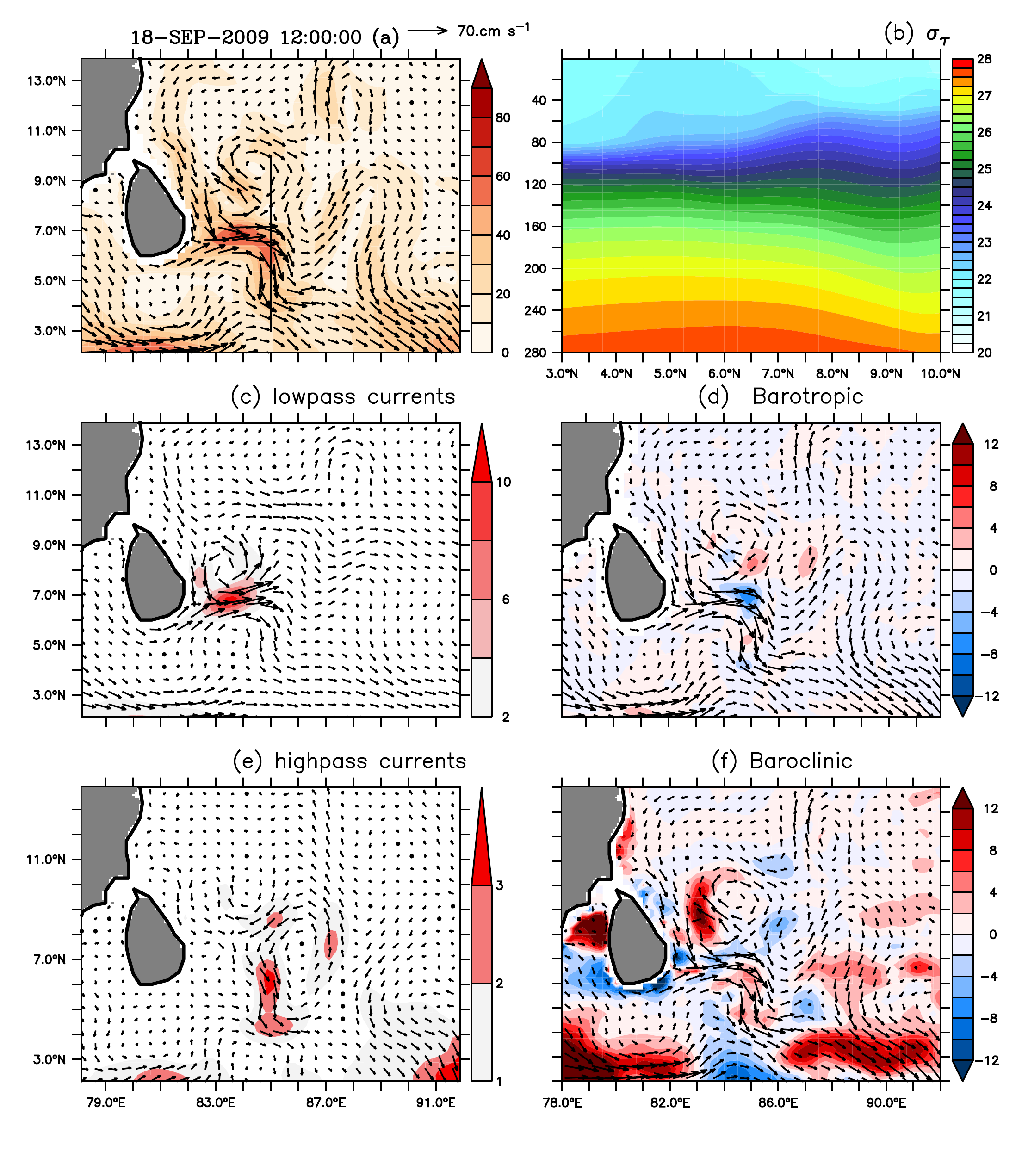}
\caption{(a) Current vectors overlaid over speed (shading, cm/s) averaged over top 30m; (b) $\sigma _{t}$ (kg/m$^{3}$); (c) 120 day lowpass filtered currents (top 30m depth averaged) as mean flow overlaid over top 100 m vertically integrated map of MKE (shading, cm$^{3}$/s$^{2}$); (d) Current vectors (top 30m depth averaged) overlaid over top 100 m vertically integrated map of barotropic energy conversion $-u_{i}^{\prime }u_{j}^{\prime }\partial \overline{u_{i}}/\partial y$ (shading, cm$^{3}$/s$^{3}$); (e) 120 day highpass filtered currents (cm/s) averaged over top 30 m, as eddies overlaid over top 100 m vertically integrated map of EKE (shading, $10^{6}$ cm$^{3}$/s $^{2}$); (f) Current vectors (top 30m depth averaged) overlaid over top 100 m vertically integrated map of baroclinic energy conversion (shading, cm$  ^{3} $/s$^{3}$). All quatities are 5--day running averaged. Vectors are of same length in all panels and the scale vector is shown in above panel (a).}
\label{fig:crmultiplot_evo85n_261}
\end{figure}

    The background flow (120-day low-pass component in our case) varied greatly
during the summer monsoon. Thus, the processes that were significant at different
phases of SMC as it evolved are presented in this section.  Evolution of SMC
is shown along with the dominant processes involved, for selected days of
the 2009 summer monsoon in Figs.\ref{fig:crmultiplot_evo85n_141}--\ref{fig:crmultiplot_evo85n_261}. Five-day running average was applied to all the
quantities to smoothen out rapid fluctuations. Evolution of SMC and the processes associated with  upper-layer currents,
their high-pass and low-pass components and analysis of energetics
are shown. The low-pass current hads a zonal orientation east of Sri Lanka at
the onset of summer monsoon in late--May. It gradually developed a
northeastward turn in mid--June, and was associated with an anticyclonic vortex
in July. The current gradually meandered in August and fed into the
southeastern BoB. The evolution of SMC viewed from
energetics perspective unfolds some important features.\bigskip
	
\subsection{Onset}
	
\begin{figure}[h]
\centering
\includegraphics[width=1\linewidth]{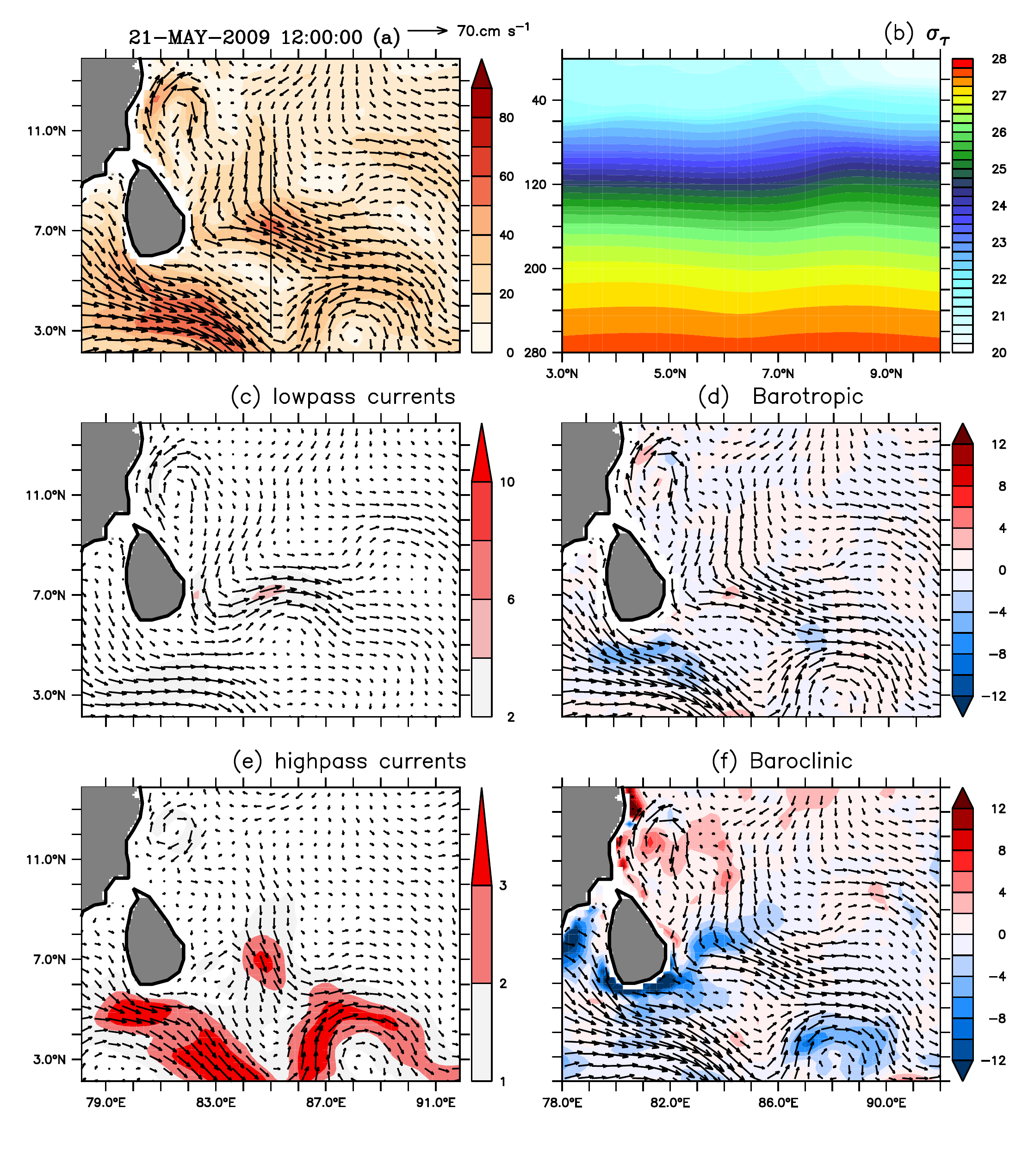}
\caption{(a) Current vectors overlaid over speed (shading, cm/s) averaged over top 30m; (b) $\sigma _{t}$ (kg/m$^{3}$); (c) 120 day lowpass filtered currents (top 30m depth averaged) as mean flow overlaid over top 100 m vertically integrated map of MKE (shading, cm$^{3}$/s$^{2}$); (d) Current vectors (top 30m depth averaged) overlaid over top 100 m vertically integrated map of barotropic energy conversion $-u_{i}^{\prime }u_{j}^{\prime }\partial \overline{u_{i}}/\partial y$ (shading, cm$^{3}$/s$^{3}$); (e) 120 day highpass filtered currents (cm/s) averaged over top 30 m, as eddies overlaid over top 100 m vertically integrated map of EKE (shading, $10^{6}$ cm$^{3}$/s $^{2}$); (f) Current vectors (top 30m depth averaged) overlaid over top 100 m vertically integrated map of baroclinic energy conversion (shading, cm$ ^{3} $/s$^{3}$). All quatities are 5--day running averaged. Vectors are of same length in all panels and the scale vector is shown in above panel (a).}
\label{fig:crmultiplot_evo85n_141}
\end{figure}

	Onset of summer monsoon in 2009 was in late--May \citep{das2015formation}.
With the onset, the wind became south-westerly and gained strength.
Circulation around Sri Lanka during this time comprised of three different
current systems: (1) a zonal flow east of Sri Lanka, (2) a southeastward
current south of Sri Lanka, and (3) a downwelling eddy (positive sea level anomalies) centered around 87$^{\circ }$E, 3$^{\circ }$N (Fig.\ \ref%
{fig:crmultiplot_evo85n_141}a). With the advent of westward propagating eddy, the southeastward current merged with the zonal current around 5$^{\circ }$N. The current was blocked by the eddy and it fed into the zonal flow east
of Sri Lanka. As the anticyclonic eddy was associated with
prominent EKE, density surfaces along 85$^{\circ }$E developed
a poleward slope with a sharp gradient between 5$^{\circ }$N and 7$^{\circ }$N (Fig.\ \ref{fig:crmultiplot_evo85n_141}b). The MKE around this time started increasing (Fig.\ \ref%
{fig:crmultiplot_evo85n_141}c). However, no notable barotropic or baroclinic
energy conversions took place during this time (Fig.\ \ref{fig:crmultiplot_evo85n_141}d,   
 \ref{fig:crmultiplot_evo85n_141}f).
\bigskip
	
	\subsection{Intensification}
	
	As the eddy moved further westward, the circulation system east of Sri Lanka was characterized by negative barotropic energy conversion (Fig.\ \ref%
{fig:crmultiplot_evo85n_150}d). This implies wave--induced accelerations to
the developing SMC, that was reflected by the intensification of surface
current, was restricted to a particular region (Figs.\ \ref%
{fig:crmultiplot_evo85n_150}a, \ref{fig:crmultiplot_evo85n_162}a). The
current intensified further in June with little mean to eddy energy
conversion noticed along 83$^{\circ }$E, as the southward flowing current
along the east coast of Sri Lanka merged with the north-eastward SMC. The 120-day high-pass filtered eddy pushed the mean flow northward. The eddies thus
seemed to impose an acceleration tendency on the SMC (Fig.\ \ref%
{fig:crmultiplot_evo85n_162}a).
\par To understand when these eddies caused the SMC to intensify, we use the TEM approach to modify the momentum and density equations. The
advantage of using TEM approach is that it allows the eddy forcing to appear
only in the horizontal momentum equation. We also use the quasi-geostrophic (QG)
approximation, so that the only eddy forcing in the TEM equation is reduced to
eddy potential vorticity (PV) flux (see Appendix B). Insights from eddy
enstrophy equation helps identify where the eddies tend to drive the mean
flow.

	\subsubsection{Where did eddies cause this intensification?}
	
	Whether the eddies tend to drive the mean flows or lead to loss of momentum
	in the mean flow depends on the nature of PV fluxes relative to the mean PV gradient. For eddies to drive mean flow, eddy PV flux has to be upwards
	relative to the mean potential vorticity gradient \citep{holland1980example}%
	, hence also known as upgradient eddy PV flux. This can be understood in
	terms of the general enstrophy equation \citep{wilson2004eddy},%
	\begin{equation}
	\frac{\partial q^{\prime 2}}{\partial t}+\boldsymbol{\nabla }\cdot 
	\boldsymbol{u}\frac{q^{\prime 2}}{2}+\boldsymbol{u}^{\prime }q^{\prime
	}\cdot \boldsymbol{\nabla }\bar{q}=-D  \label{eqn:q'(2)}
	\end{equation}%
	where $\overline{q}$ is the mean PV, $q^{\prime }$ is the eddy PV and 
	\textit{D} represents the dissipation of eddy enstrophy, all other symbols
	bearing their usual meanings as discussed earlier. A positive dissipation
	would lead to eddy enstrophy decay. First term on the left hand side of the
	equation (\ref{eqn:q'(2)}) represents the local rate of change of eddy
	enstrophy, second term gives the divergence of eddy enstrophy advection and
	third term however gives us the eddy enstrophy decay termed here as \textit{%
		eddy destruction}. $\boldsymbol{\nabla }\cdot \boldsymbol{u}\left( q^{\prime 2}/2\right) $ is termed as  ``enstrophy destruction'' term as the latter dissipates the eddies by fluxing PV into the mean flow. Negative of $\boldsymbol{\nabla }\cdot \boldsymbol{u}\left( q^{\prime 2}/2\right) $ can be thought of as
	generation of eddies as it leads to local increment in eddy enstrophy in (%
	\ref{eqn:q'(2)}).\\ For the eddies to drive mean flow there has to be a decay
	of eddy enstrophy in the vicinity of eddy mean flow interaction. The second
	term can be considered as a higher order term in the equation which helps us
	to identify regions of convergence and divergence of eddy enstrophy flux.
	 
\begin{figure}
\centering
\includegraphics[width=\linewidth]{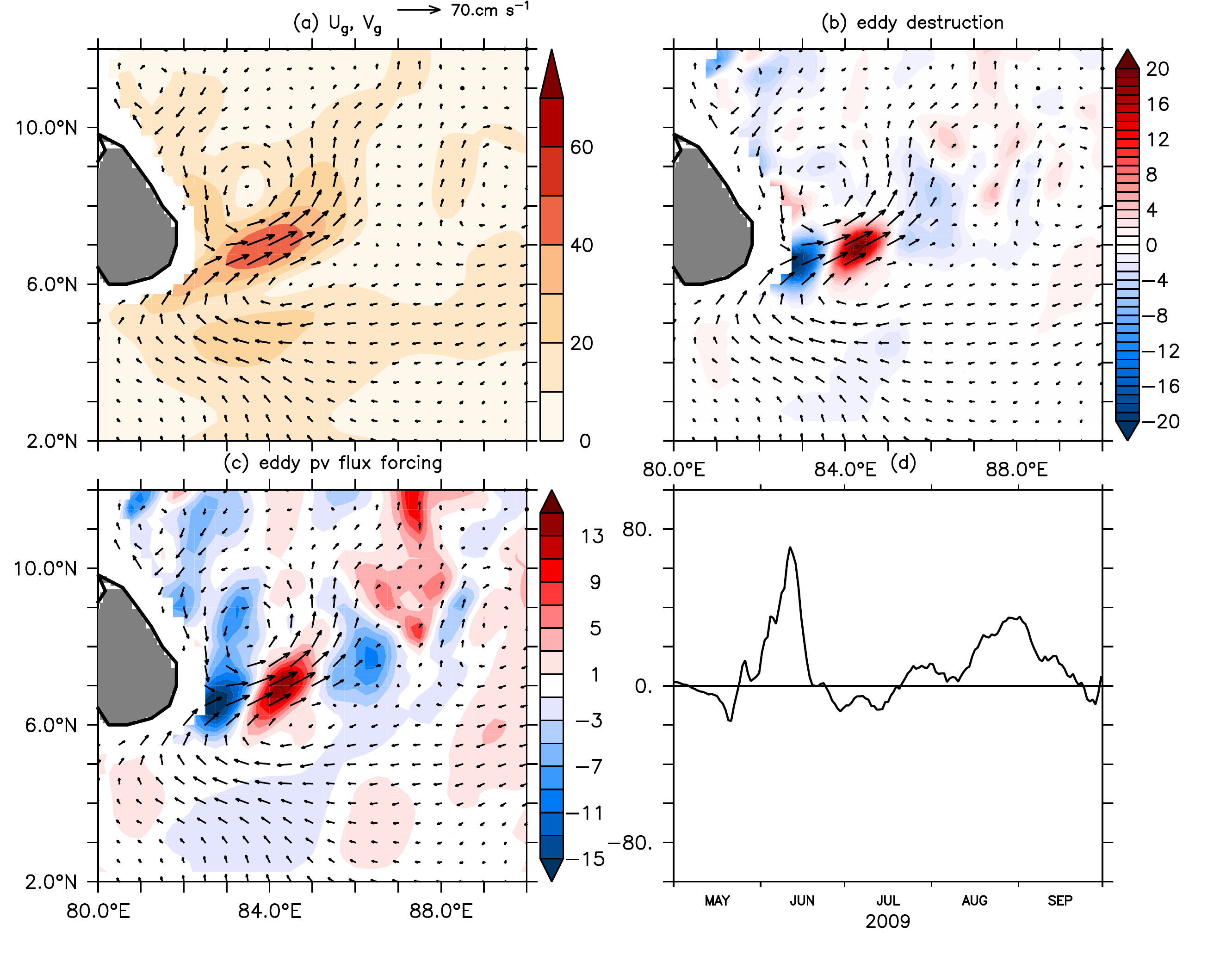}
\caption{(Clockwise) At 50 m, time averaged (June-September) geostrophic currents vectors overlaid over speeds (shading, cm/s); geostrophic current vectors overlaid over eddy enstrophy destruction term $\textbf{u$'$}q'\cdot\nabla \overline{q}$ (shading, $10^{-16} cm/s^2$); time series of eddy potential vorticity flux forcing, $v^{\prime
		}q^{\prime }$ ($10^{-5}$ cm/s$^{2}$) at 84$^{\circ }$E, 6.5$^{\circ }$N
		and at 50 m depth; and eddy pv flux \textit{$v'q'$}(shading, $10^{-5} cm/s^2$). Vectors are of the same scale as in Panel 1 (top left).}
\label{fig:paper_edd_mean_flow_summer50}
\end{figure}

	Insights from the eddy destruction (Fig.\ \ref{fig:paper_edd_mean_flow_summer50}b) help
	to identify regions of upgradient eddy PV flux. Here the eddies try to drive
	the mean flow by accelerating the latter. Regions of high velocities in the
	SMC are found to be associated with upgradient eddy PV flux. Upstream the
	eddy PV flux is downgradient indicating regions where eddies extract
	momentum and energy from mean flow. The region is also
	characterised by positive eddy PV flux downstream and
	negative upstream
	(Fig.\ \ref{fig:paper_edd_mean_flow_summer50}c). This implies that the eddies extract
	momentum and energy from the SMC upstream. The same region is also
	barotropically unstable as found from energy considerations (Fig.\ \ref{fig:paper_energy_conv}).
	
		\subsubsection{When did eddies cause this intensification?}
		
	Time series of eddy PV flux, $v^{\prime}q^{\prime}$ 
	provides a more comprehensive view of the eddy mean flow interaction. Regions where the eddies tended to drive SMC are identified (Fig.\ \ref{fig:paper_edd_mean_flow_summer50}b) and a location has been chosen from the region to identify when the eddies caused an acceleration tendency on the mean flow.  The time series of eddy PV flux at 84$^{\circ }$E, 6.5$^{\circ }$N and
at 50--m depth indicates a sharp positive peak in mid--June (Fig.\ \ref%
{fig:paper_edd_mean_flow_summer50}d). This coincided with the arrival of the
Rossby waves and westward propagating eddies in the region. Thus, the eddies,
because of positive values of $v^{\prime}q^{\prime}$ (see Appendix B, C) could be
expected to provide an acceleration tendency to the mean flow as they arrived
and interacted with the latter in mid-June. Throughout July, the eddy PV flux
remained relatively low at the location. Positive eddy PV flux was observed
in late--August and early--September, but by this time, the SMC already started to
weaken and disappear.
	
\subsubsection{Local winds}

\begin{figure}[h]
\centering
\includegraphics[width=1\linewidth]{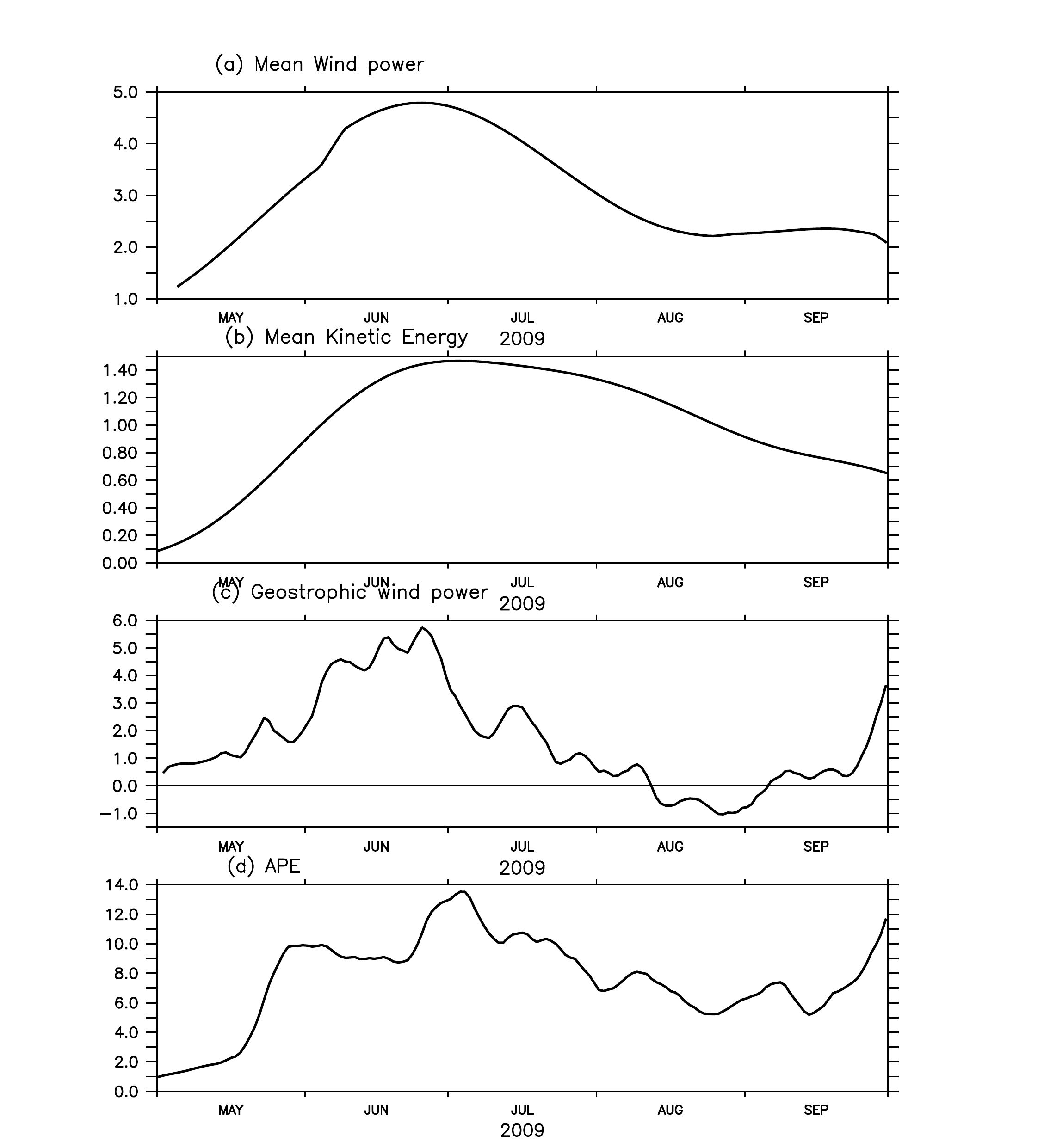}
\caption{(a) Time series of area integrated (82--86$^{\circ }$E, 5--8$^{\circ }$ N) mean wind Power (5-day smoothened) (top) ($10^{9}$ W) (b) surface integrated (82--86$^{\circ }$E, 5--8$^{\circ }$N) MKE averaged over top 30 m (5-day smoothened) ($10^{14}$ Joules/s$^{2}$); (c) Area integrated (82--86$ ^{\circ }$E, 5--8$^{\circ }$N) geostrophic wind power (5-day smoothened) (top) ($10^{9}$ W); (d)Volume integrated (82--86$^{\circ }$E, 5--8$^{\circ }$ N, 0--100 m) APE (5-day smoothened) ($10^{14}$ Joules)}
\label{fig:paper_eddy_wind}
\end{figure}

However, the local winds also became very strong over the region during this
time, and thus could influence the MKE significantly (Fig.\ \ref%
{fig:crmultiplot_evo85n_150}c). The MKE of the ocean
current is related to the mean wind power in 
equation (2) as 
\begin{equation}
\int_V \rho _{0}\frac{\partial {\overline{\boldsymbol{u}^{2}}}}{\partial t}
\, dV \sim \int_A\overline{\mathcal{F}\cdot {u}}\, dA = \int_A\overline{%
\boldsymbol{\tau }\cdot \boldsymbol{u}}\,dA  \label{eqn:KE&windforcing}
\end{equation}
where $\tau$ is the wind stress. Figures \ref{fig:paper_eddy_wind}a and %
\ref{fig:paper_eddy_wind}b shows the variation of mean wind power input
along with MKE, integrated over the top 40 m within
the surface mixed layer. Both quantities are integrated over 82--86$%
^{\circ }$E, 5--8$^{\circ }$ N. The MKE increased with
the increase in total mean wind power in June, peaked in late-June following the
variation of mean wind power, and started to decline from mid- July. Thus, the
wind can be considered as the predominant driving agent in intensifying
mean SMC flow.

	\subsection{Anticyclonic bend}
	
\begin{figure}[h]
\centering
\includegraphics[width=1\linewidth]{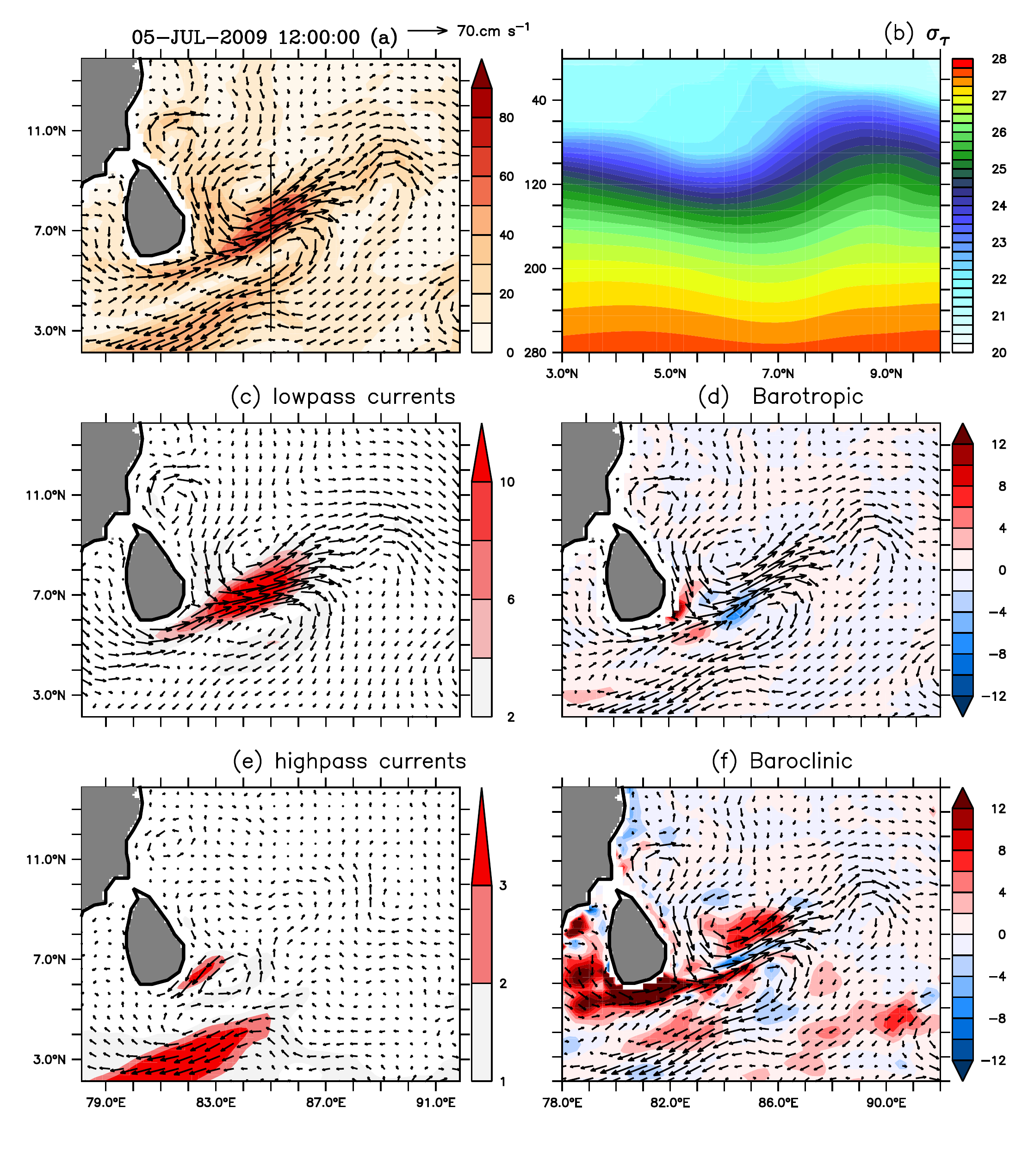}
\caption{(a) Current vectors overlaid over speed (shading, cm/s) averaged over top 30m; (b) $\sigma _{t}$ (kg/m$^{3}$); (c) 120 day lowpass filtered currents (top 30m depth averaged) as mean flow overlaid over top 100 m vertically integrated map of MKE (shading, cm$^{3}$/s$^{2}$); (d) Current vectors (top 30m depth averaged) overlaid over top 100 m vertically integrated map of barotropic energy conversion $-u_{i}^{\prime }u_{j}^{\prime }\partial \overline{u_{i}}/\partial y$ (shading, cm$^{3}$/s$^{3}$); (e) 120 day highpass filtered currents (cm/s) averaged over top 30 m, as eddies overlaid over top 100 m vertically integrated map of EKE (shading, $10^{6}$ cm$^{3}$/s $^{2}$); (f) Current vectors (top 30m depth averaged) overlaid over top 100 m vertically integrated map of baroclinic energy conversion (shading, cm$ ^{3} $/s$^{3}$). All quatities are 5--day running averaged. Vectors are of same length in all panels and the scale vector is shown in above panel (a).}
\label{fig:crmultiplot_evo85n_186}
\end{figure}

	At the beginning of July, part of the SMC started to take an anticyclonic turn
near 85$^{\circ }$E, while the other part continued to flow northeastward
(Fig.\ \ref{fig:crmultiplot_evo85n_186}a, \ref{fig:crmultiplot_evo85n_186}%
c). The MKE was high in the region during this time (Fig.\ %
\ref{fig:crmultiplot_evo85n_186}c), while the EKE was
significantly low (Fig. \ref{fig:crmultiplot_evo85n_186}e). Barotropic
energy conversion was also very small (Fig.\ref%
{fig:crmultiplot_evo85n_186}d). However, some baroclinic energy
conversions took place north of the SMC, where it took  cyclonic turn
(Fig. \ref{fig:crmultiplot_evo85n_186}f). Density surfaces within 40--100 m
along 85$^{\circ }$E were suppressed between 5$^{\circ }$N and 7$^{\circ }$%
N, with a poleward gradient, which resulted in the generation of APE. In mid--July, the anticyclonic turn of the SMC became more
prominent associated with high MKE but low EKE. Barotropic energy conversion was very low in the region, while
positive baroclinic energy conversion existed along the SMC. Negative
baroclinic energy conversion implying an increase in eddy available
potential energy can be found on both sides of the SMC. But, this does not
adequately explain the bending of the SMC. The anticyclonic turning of the
low-pass current could also be due to the local winds in the region. The
ageostrophic component of the current supported this possibility (Fig.\ \ref%
{fig:ageos}). The ageostrophic currents were larger and modulated the
background flow in the absence of prominent eddies in the region during
the time. The baroclinic energy conversion observed could be due to the
variation in wind energy input, leading to the fluctuation of the
density surfaces (to be discussed next).\bigskip 
%	Poleward gradient of the density surfaces (Fig.\ \ref{fig9:currents&sigmaT(jun20)}b) continue to
%	increase probably owing to the the wind work being utilised to raise the
%	isopycnals \citep{huang2010ocean}.\bigskip 
  
   \subsection{Anticyclonic vortices}

\begin{figure}[h]
\centering
\includegraphics[width=1\linewidth]{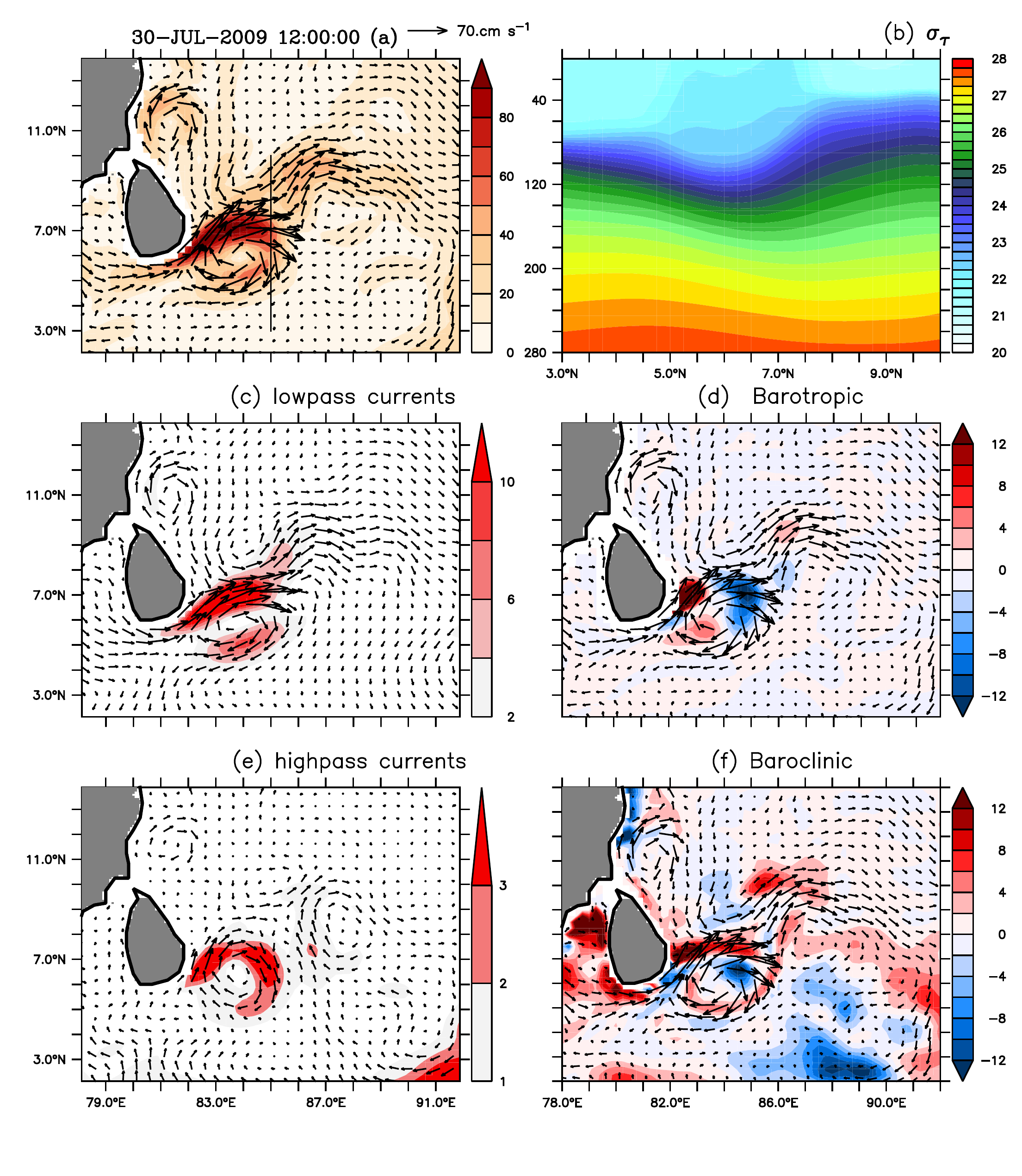}
\caption{(a) Current vectors overlaid over speed (shading, cm/s) averaged over top 30m; (b) $\sigma _{t}$ (kg/m$^{3}$); (c) 120 day lowpass filtered currents (top 30m depth averaged) as mean flow overlaid over top 100 m vertically integrated map of MKE (shading, cm$^{3}$/s$^{2}$); (d) Current vectors (top 30m depth averaged) overlaid over top 100 m vertically integrated map of barotropic energy conversion $-u_{i}^{\prime }u_{j}^{\prime }\partial \overline{u_{i}}/\partial y$ (shading, cm$^{3}$/s$^{3}$); (e) 120 day highpass filtered currents (cm/s) averaged over top 30 m, as eddies overlaid over top 100 m vertically integrated map of EKE (shading, $10^{6}$ cm$^{3}$/s $^{2}$); (f) Current vectors (top 30m depth averaged) overlaid over top 100 m vertically integrated map of baroclinic energy conversion (shading, cm$ ^{3} $/s$^{3}$). All quatities are 5--day running averaged. Vectors are of same length in all panels and the scale vector is shown in above panel (a).}
\label{fig:crmultiplot_evo85n_211}
\end{figure}

    In late--July (Fig.\ \ref{fig:crmultiplot_evo85n_211}a), an
anticyclonic vortex was fully developed and centered around 84$^{\circ }$E. In
addition to the anticyclonic turning of the low-pass background current, an
anticyclonic eddy associated with strong EKE appeared in
this region. Both barotropic and baroclinic energy conversions were
responsible for the generation of these eddies.. The SMC at 83$^{\circ }$E, 6$^{\circ }$N became barotropically unstable, leading to the conversion MKE to EKE (Fig.\ \ref{fig:crmultiplot_evo85n_211}d).
Baroclinic instability also appeared around 7$^{\circ }$N associated with
flattened isopycnals that weakened the flow and could cause the current to
meander (Fig.\ \ref{fig:crmultiplot_evo85n_211}f). Significant baroclinic
energy conversion appeared all along the outer periphery of the
anticyclonic vortex. In mid--July, at around 89$^{\circ }$E, 8$^{\circ }$N, the SMC encountered another
anticyclonic eddy, which remained non-interacting in terms of energy exchanges. Later in late--July, due to
the gradual meandering pattern of the current, the eddy around 8$^{\circ }$N
further intruded into the SMC (Fig.\ \ref{fig:crmultiplot_evo85n_211}a).
This eddy was termed as an Intermediate Geostrophic (IG) eddy %
\citep{vinayachandran1998monsoon}.\bigskip

   \subsection{Meandering}

Horizontal shear of the SMC was reduced as a result of barotropic instability.
Further, in early--August (Fig.\ \ref{fig:crmultiplot_evo85n_231}b), the
sloping density surfaces flattened, releasing a large amount of APE. Baroclinic instability associated with this
flattened density gradients and meridional buoyancy flux was the dominant
mechanism leading to the growth of disturbances (Fig. \ref%
{fig:crmultiplot_evo85n_231}f). The combined effect of barotropic and
baroclinic instabilities weakened the SMC which now meandered because of
baroclinic instability \citep{ikeda1981meanders}. One possible reason behind
the release of APE in late--July and August could be the gradual weakening of local wind power.
	
	\subsubsection{Local wind power}
	
	The wind power is calculated by
\begin{equation}
\nonumber 
	W=\iint\limits_{A}\boldsymbol{\tau \cdot u}\,dA, =\iint\limits_{A}(\boldsymbol{\tau \cdot u_g}+ \boldsymbol{\tau \cdot u_{ag}})\,dA,
	\end{equation}% }

where $\boldsymbol{\tau \cdot u_g}$ is the geostrophic wind power while $\boldsymbol{\tau \cdot u_{ag}}$ is the ageostrophic wind power. The geostrophic wind work is utilized in vertically displacing the
	isopycnals thereby influencing APE %
	\citep{brown2010much} via Ekman pumping, upwelling and downwelling. On the
	other hand, the ageostrophic wind work is required to maintain the Ekman
	spiral \citep{huang2010ocean}.
	
\begin{figure}[h]
\centering
\includegraphics[width=1\linewidth]{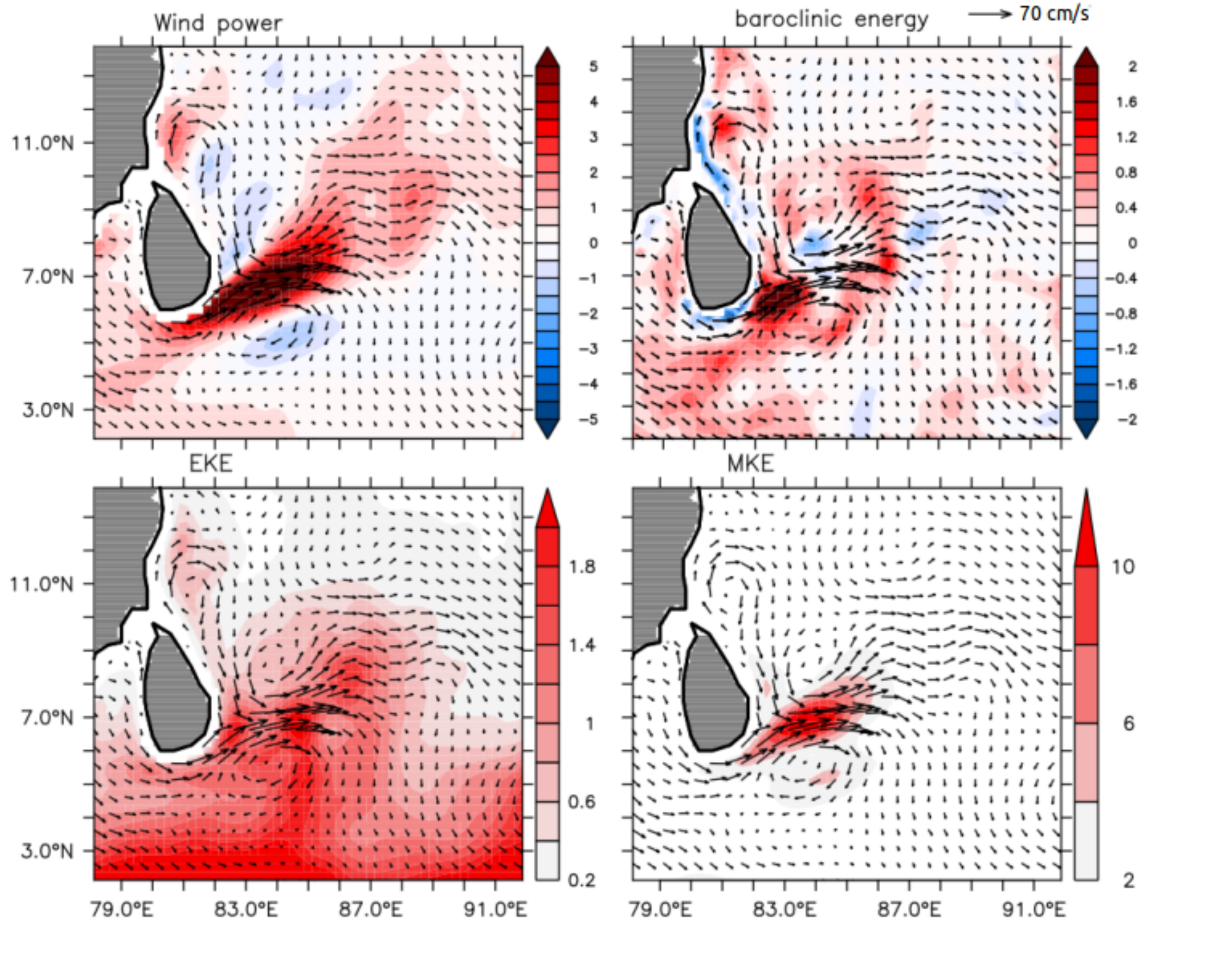}
\caption{Clockwise: Time mean (June--September) maps of current vectors (averaged over top 30 m) overlaid over wind power (shading, watts); current vectors overlaid over vertically integrated (top 100 m) baroclinic energy conversions (shading, cm$^{3}$/s$^{3}$) ; current vectors overlaid over vertically integrated (top 100 m) mean kinetic energy and eddy kinetic energy (shading, cm$^{3}$/s$^{2}$).}
\label{fig:wind_eddy_map_sumr1_main_current}
\end{figure}

	Figure \ref{fig:wind_eddy_map_sumr1_main_current} shows time-mean (June--September) maps of total wind power (combining geostrophic and ageostrophic),
	baroclinic energy conversions, MKE and EKE, vertically integrated over 100m depth. The wind power, MKE and EKE together with baroclinic energy
	conversions are also found to be significantly high over this region.
	
	\subsubsection{Available potential energy}

APE can be defined as difference between total
	potential energy and minimum total potential energy which could result from
	any adiabatic redistribution of mass. It can be regarded as a deviation from
	the resting (minimum energy configuration) stable stratification state. To estimate
	local APE, we calculate it in Joules as \citet{oort1989new} 
	\begin{equation}
	E_{p}=-\frac{g}{2}\iiint\limits_{V}\frac{(\rho -\overline{\rho })^{2}}{%
		\partial \overline{\rho }/\delta z}dV.  \label{eqn:APE}
	\end{equation}%
	%
	%
	%where N is buoyancy frequency.
	%\begin{linenomath*}
	%\begin{equation}
	%N^2=-\frac{g\frac{\partial\rho_r}{\partial z}}{\rho_0}
	%\end{equation}
	%\end{linenomath*}
	%\noindent Since the stratification is not linear so higher order terms in estimating the available potential energy were required. 
	APE has been integrated over 82--86$^{\circ }$E, 5--8$^{\circ }$N and from
	the surface to 100 m. %\begin{figure}[t!]
	
	Figures \ref{fig:paper_eddy_wind}c and \ref{fig:paper_eddy_wind}d shows the relationship between wind
	power and volume integrated APE. The APE corresponded well with the variation in wind power. Only in
	the month of June, the APE remained unchanged for some
	time inspite of increasing wind power. The arrival of downwelling Rossby waves
	associated with suppressed thermocline could negate the influence of wind
	power. In late--June, APE again started to
respond to change in wind power. During this time, the local wind power
probably overwhelmed the effects of westward propagating eddies. Significant amount
of APE reduction started from mid--July and continued till August, which was associated with the possibility of major baroclinic instability taking place at the depth.
	
	\subsubsection{Westward propagating eddies}
	
\begin{figure}[h]
\centering
\includegraphics[width=1\linewidth]{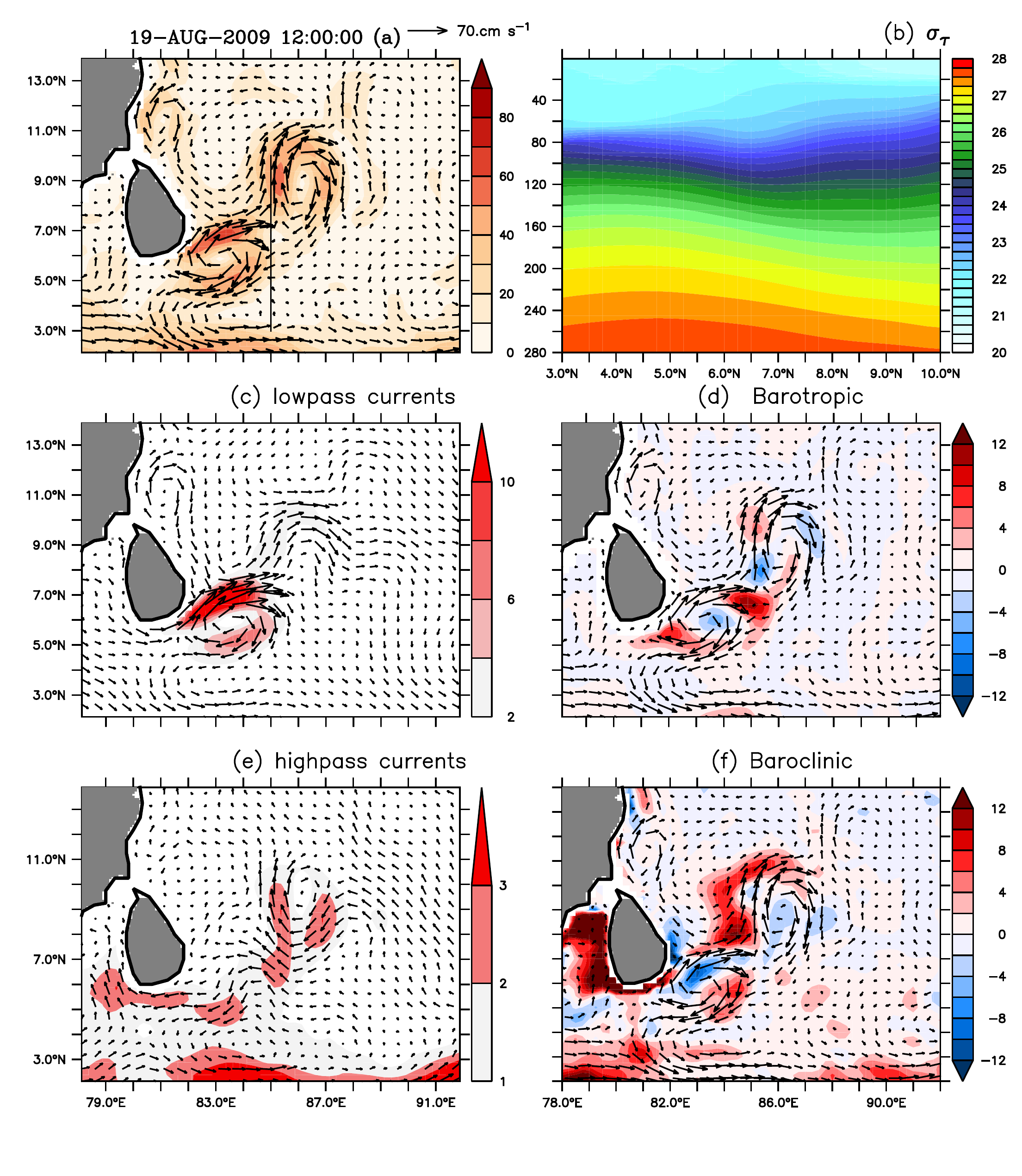}
\caption{(a) Current vectors overlaid over speed (shading, cm/s) averaged over top 30m; (b) $\sigma _{t}$ (kg/m$^{3}$); (c) 120 day lowpass filtered currents (top 30m depth averaged) as mean flow overlaid over top 100 m vertically integrated map of MKE (shading, cm$^{3}$/s$^{2}$); (d) Current vectors (top 30m depth averaged) overlaid over top 100 m vertically integrated map of barotropic energy conversion $-u_{i}^{\prime }u_{j}^{\prime }\partial \overline{u_{i}}/\partial y$ (shading, cm$^{3}$/s$^{3}$); (e) 120 day highpass filtered currents (cm/s) averaged over top 30 m, as eddies overlaid over top 100 m vertically integrated map of EKE (shading, $10^{6}$ cm$^{3}$/s $^{2}$); (f) Current vectors (top 30m depth averaged) overlaid over top 100 m vertically integrated map of baroclinic energy conversion (shading, cm$ ^{3} $/s$^{3}$). All quatities are 5--day running averaged. Vectors are of same length in all panels and the scale vector is shown in above panel (a).}
\label{fig:crmultiplot_evo85n_231}
\end{figure}

	Before this stage, the westward propagating eddy at 8$^{\circ }$N that did not interact
with the SMC (Figs.\ \ref{fig:crmultiplot_evo85n_211}d, \ref%
{fig:crmultiplot_evo85n_211}f), started exchanging energy with
meandering current (Figs. \ref{fig:crmultiplot_evo85n_231}d, %
\ref{fig:crmultiplot_evo85n_231}f). One possible reason could be the weakening
of the SMC as a result of baroclinic instability, which now reached the {critical
level} (a level where exchange of energy takes place between mean flow and
wave). The interaction of this eddy with the SMC lead to barotropic energy conversions
at 85$^{\circ }$E, 6$^{\circ }$ N (Fig. \ref{fig:crmultiplot_evo85n_231}d). This rendered the
currrent barotropically unstable and merely amplified the recirculation at 8$%
^{\circ }$N, and the current was thus found to meander and feed into the southeastern BoB (Fig. \ref{fig:crmultiplot_evo85n_231}a).

	\subsection{Termination}
	
	In late August and September, the continuous SMC disappeared (Fig.\ %
	\ref{fig:crmultiplot_evo85n_261}a). The anticyclonic eddy near 8$^{\circ
	}$N pushed against the weakened SMC and moved it westward. The entire meandering
	pattern shifted slightly to the west and by mid-September (Fig.\ \ref%
	{fig:crmultiplot_evo85n_261}a) the SMC system disappeared.
	
\clearpage
	\section{Summary and conclusions\label{summary&conclusions}}
	
	The SMC around Sri Lanka plays a crucial role in the
interaction of the water masses between the two basins on either side of
India, namely, the Arabian Sea  and Bay of Bengal. Past studies showed that the SMC is highly energetic
and eddy activity is intense in this region (\citealp{vinayachandran1998monsoon}, \citealp{shenoi1999near}, %
\citealp{chen2012features}). The prime objective of our analysis was to
understand the evolution, including the intensification and meandering of the SMC, using an
OGCM (MOM4p1).  We made several new contributions toward understanding the
evolution of SMC around Sri Lanka, which involve eddy--mean flow
interactions and effect of local versus remote forcing on the dynamics in
the region. Analysis of eddy potential vorticity eddy potential vorticity flux, eddy entrophy,
eddy energy budgets were carried out which shed new insights on how
the eddies drove the mean flow and vice versa. The study also provided
better understanding of the influence of stratification in terms of driving the mean
flow and isolating the role of winds in affecting the circulation. The evolution of SMC was classified into six stages, namely, {onset}, {intensification}, {anticyclonic bend}, {anticyclonic vortices}, {meandering}, and {termination}. Barotropic and baroclinic energy conversions were the dominant
mechanisms to modulate the background flow of the SMC. 
\par The arrival of Rossby waves and westward propagating eddies into this region
resulted in high eddy momentum flux exchange in the month of June.
Using the TEM approach and combining the net effect of
momentum and buoyancy fluxes, we showed that the eddies imparted PV flux and induced acceleration tendency on the
mean flow in mid--June. Using quasi-geostrophic approximation to
understand the effect of eddies on the mean flow, we found that there were
regions of up-gradient eddy PV flux where the eddies drove the mean
flow. There were also upstream regions associated with down-gradient eddy PV
flux where the eddies extracted energy and momentum out of the mean flow.
Local winds affected both the mean flow of the SMC and its course path during the
summer monsoon. The volume-integrated MKE was at least twice
that of EKE, suggesting that the circulation in this
region was largely wind driven. Also, the time-mean (June--September) maps of mean kinetic
MKE corresponded well with the wind power. The time series of 
ocean kinetic energy, averaged within mixed layer depth of 40 m, revealed strong correspondence with the time series of
area-integrated mean wind power, implying the dominant influence of
local winds in driving the mean current. The geostrophic wind power also
modulated the density surfaces, thereby influencing the APE except in mid--June. One explanation for such behaviour can be the arrival of downwelling
Rossby waves associated with suppressed thermocline and thereby trying to
negate the effects of winds. The variation in wind power led to generation
of eddies in the region through baroclinic energy conversion (from APE to EKE). 
\par 
It may be noted that the analyses presented here have certain limitations. The analysis in this study were
carried out for only one year. Even though this general behavior of the SMC is
similar every year, there can be considerable variability from year to year.
Therefore, it is useful to carry out the analyses for a longer period using simple linear models as nonlinearities involved in our model may obscure several important features.  
	
	\section*{Acknowledgments}
	
This research was supported by Ministry of Human Resource Development (MHRD)
 funding. OSCAR data provided by NOAA-NASA ( http://dx.doi.org/10.5067/
 OSCAR-03D01), QuikSCAT winds by IFREMER (http://www.ifremer.fr/cersat/en/data/
 data.htm) were used for the study. We sincerely thank Prof.J. P. McCreary for the insightful discussions. We also thank Dr. Zuojun Yu for her contributions in the work. Computations were carried out on a high 
 performance computer (HPC) system facility funded by Department of Science and      
 Technology under Fund for Improvement of ST Infrastructure in Universities and 
 Higher Educational Institutions (FIST) scheme at Center for Atmospheric and 
 Oceanic Sciences (CAOS), Indian Institute of Science, Bangalore and Divecha
 Center for Climate Change (DCCC). The authors are thankful to GFDL for providing 
 the MOM4p1 source code. Ferret and Matlab has been used for data analysis and 
 graphics output. 
    	
    	\pagebreak
%\appendix
\begin{appendices}

\section{Background state}
\label{appendix:A}

Let $q$ be a model variable. Then, a key aspect of all our eddy analyses is
the separation of $q$ into mean (or slowly- varying)\ $\bar{q}$ and
time-varying $q^{\prime }$ ($q=\bar{q}+q^{\prime }$). Ideally, the
mean (background) state should be free of high-frequency disturbances, that
is, free from the influence of any eddies. We define $\bar{q}$ to be the
model field during 2009 smoothed by a 120-day, Lanczos low--pass filter, and $%
q^{\prime }$ to be the field obtained using a 120-day high--pass filter. We
tried several cutoff periods, finding 120 days optimal: it is well
above the residence period of mesoscale eddies in the region, and at the
same time it retains the large-scale seasonal variability. The background state
is the same for both eddy energy budget and TEM analysis.

    	\section{TEM equations}
%\label{appendix:B}

\subsubsection{Equations of motion}
	
	The horizontal momentum and buoyancy equations can be decomposed into a mean
	and perturbation part as shown in (1). These equations under
	quasigeostrophic (QG) approximation in Reynold's averaged form become%
	\begin{equation}
	\frac{\partial \overline{\boldsymbol{u}_{g}}}{\partial t}+\overline{%
		\boldsymbol{u}_{g}}\cdot \nabla \overline{\boldsymbol{u}_{g}}+f_{0}%
	\boldsymbol{k}\times \overline{\boldsymbol{u}_{ag}}+\beta y\boldsymbol{k}%
	\times \overline{\boldsymbol{u}_{g}}-\frac{\boldsymbol{\nabla }\overline{{%
				p_{ag}}}}{\rho _{0}}=\boldsymbol{G}-\boldsymbol{\nabla }\cdot \overline{%
		\boldsymbol{u}^{\prime }\boldsymbol{u}^{\prime }},  \label{eqn:ugbar(1)}
	\end{equation}%
	\begin{equation}
	\boldsymbol{\nabla }_{h}\cdot \overline{\boldsymbol{u}_{ag}}+\frac{\partial 
		\overline{w_{ag}}}{\partial z}=0,  \label{eqn:divuagbar}
	\end{equation}%
	\begin{equation}
	\frac{\partial \overline{b}}{\partial t}+\overline{\boldsymbol{u}_{g}}\cdot 
	\boldsymbol{\nabla }\overline{b}+\overline{w_{ag}}{N_{0}^{2}}=X-\boldsymbol{%
		\nabla }\cdot \overline{\boldsymbol{u}^{\prime }b^{\prime }},
	\label{eqn:bbar(1)}
	\end{equation}%
	where ${u_{g}}$ represents geostrophic velocities, the overbar denotes the
	120-day, low-pass-filtered background state and the primed quantities
	represent high-pass-filtered eddies. $u_{ag}$, $w_{ag}$ are the ageostrophic
	components, $b$ represents the buoyancy and $G$, $X$ are forcing terms. Thus
	the effect of eddies on background state appear in two equations (Eqs.\ \ref%
	{eqn:ugbar(1)} and \ref{eqn:bbar(1)}) represented by the convergence of eddy
	momentum fluxes, $-\boldsymbol{\nabla }\cdot \overline{\boldsymbol{u}%
		^{\prime }\boldsymbol{u}^{\prime }}$ and buoyancy flux, $-\boldsymbol{\nabla 
	}\cdot \overline{\boldsymbol{u}^{\prime }b^{\prime }}$.
	
	Apart from having an acceleration tendency on background flow, the momentum
	flux can also drive the horizontal mean ageostrophic flow. Also, the
	buoyancy fluxes can drive the mean flow implying the dynamical coupling
	between density and momentum equations. It is not thus possible to view how
	the density fluxes may accelerate or decelerate the background flow.
	Transformed Eulerian mean (TEM) approach helps us to get rid of this problem
	as a result of which only the eddy forcing appears in the horizontal
	momentum equations. The TEM equations can be written as%
	\begin{equation}
	\frac{\partial \overline{\boldsymbol{u}_{g}}}{\partial t}+\overline{%
		\boldsymbol{u}_{g}}\cdot \boldsymbol{\nabla }\overline{\boldsymbol{u}_{g}}%
	+f_{0}\boldsymbol{k}\times \widetilde{\boldsymbol{u}_{ag}}+\beta y%
	\boldsymbol{k}\times \overline{\boldsymbol{u}_{g}}-\frac{\boldsymbol{\nabla }%
		\overline{{p_{ag}}}}{\rho _{0}}=\boldsymbol{G}-\underbrace{\boldsymbol{%
			\nabla }\cdot \boldsymbol{E}}_{EF},  \label{eqn:ugbar(2)}
	\end{equation}%
	\begin{equation}
	\boldsymbol{\nabla }_{h}\cdot \widetilde{\boldsymbol{u}_{ag}}+\frac{\partial 
		\widetilde{w_{ag}}}{\partial z}=0,  \label{eqn:divuagtilde}
	\end{equation}%
	\begin{equation}
	\frac{\partial \overline{b}}{\partial t}+\overline{\boldsymbol{u}_{g}}\cdot 
	\boldsymbol{\nabla }\overline{b}+\widetilde{w_{ag}}{N_{0}^{2}}=X
	\label{eqn:bbar(2)}
	\end{equation}%
	where $EF$ = eddy forcing. The tilde over the ageostropic circulation terms
	represents the residual mean circulation. In many cases, the mean
	ageostrophic circulation can negate the impact of eddies. The residual mean
	circulation represents the net effect of the two quantities. Eddy components
	are given by the eddy bolus velocities,%
	\begin{equation}
	\boldsymbol{u}^{\mathbf{\ast }}=-\frac{\partial }{\partial z}\frac{\overline{%
			b^{\prime }\boldsymbol{u}\mathbf{^{\prime }}}}{N_{0}^{2}},\hspace{2em}%
	\boldsymbol{w}^{\mathbf{\ast }}=\boldsymbol{\nabla }\cdot \frac{\overline{%
			b^{\prime }\boldsymbol{u}\mathbf{^{\prime }}}}{N_{0}^{2}}.
	\label{eqn:bolusvelocities}
	\end{equation}
	
	\section{Eddy potential vorticity flux\label{EPVflux}}
	
	Vector $\boldsymbol{E}$ in (\ref{eqn:ugbar(2)}) represents the eddy-stress
	tensor \citep{plumb1986three}. The advantage of using the QG approximation
	to understand eddy-driven mean flows is that it expresses the divergence of
	eddy stress tensor explicitly in terms of eddy potential-vorticity flux. 
	\begin{equation}
	\boldsymbol{\nabla }\cdot \boldsymbol{E}\approx -\boldsymbol{\hat{k}}\times 
	\overline{q^{\prime }\boldsymbol{u}^{\prime }},  \label{eqn:divE}
	\end{equation}%
	where%
	\begin{equation}
	{q^{\prime }}\boldsymbol{u}{^{\prime }}=\boldsymbol{\nabla }\cdot \left( 
	\begin{array}{ccc}
	N & M-P & 0 \\ 
	M-P & -N & 0 \\ 
	R & S & 0%
	\end{array}%\counterwithout{equation}{section}
	\right) ,  \label{eqn:q'u'vector}
	\end{equation}%
	$M=\left( \overline{v^{\prime 2}-u^{\prime 2}}\right) /2$, $N=\overline{%
		u^{\prime }v^{\prime }}$, $P=\left( \overline{b^{\prime 2}}/2\right) N^{2}$, 
	$R=\overline{v^{\prime }b^{\prime }}$, and $S=\left( f_{0}/N_{0}^{2}\right) 
	\overline{u^{\prime }b^{\prime }}$. Variables $M$ and $N$ are the eddy
	momentum flux, $P$ is the eddy potential energy, and $R$ and $S$ are the
	meridional and zonal eddy buoyancy fluxes under the QG assumption. Although
	both eddy momentum and buoyancy fluxes impact the mean flow, considering
	their effects separately can sometimes be misleading. This problem is
	avoided in $\boldsymbol{E}$, which combines the influence of the two fluxes.
	The component of $\boldsymbol{E}$ normal to the mean flow tends to
	accelerate or decelerate it, whereas the component parallel to the mean flow
	tends to turn it.
\end{appendices}
    	\clearpage

	\clearpage
	%EndExpansion
%\renewcommand\thesection{}
	%\section{References\protect\vspace{-0.25in}}

\bibliographystyle{apalike}
\bibliography{author}
\clearpage
%1
\clearpage
%2
\begin{figure}[h]
\centering
\includegraphics[width=1\linewidth]{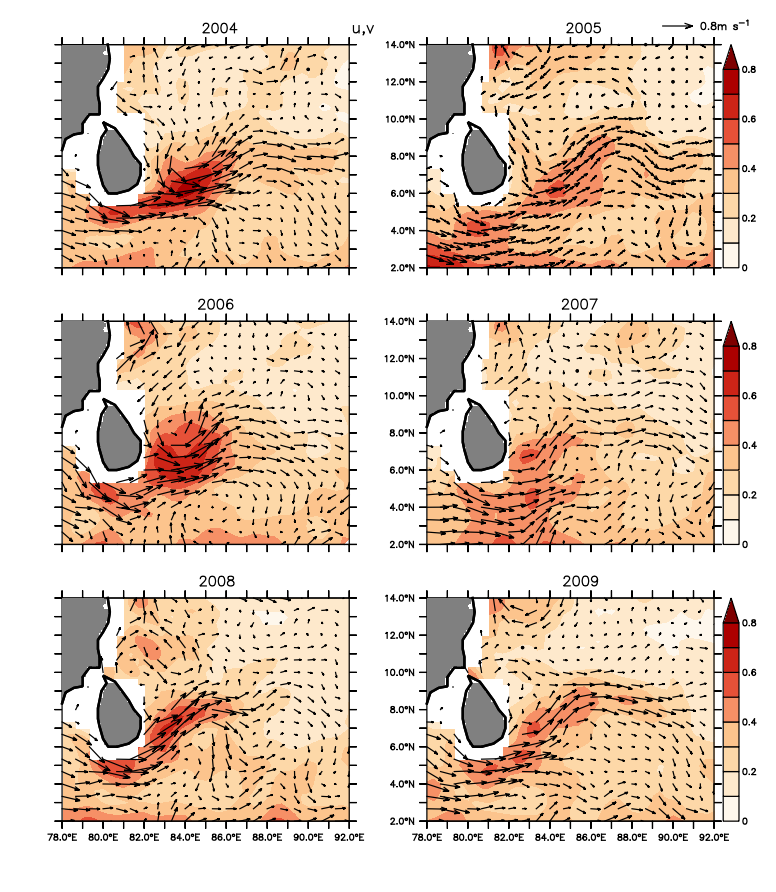}
\caption{Time mean (June-- September) circulation map of SMC from 2004--2009. The scale vector (0.8 m/s) is shown above panel.}
\label{fig:inter_annual_smc1}
\end{figure}
\clearpage
%3

\clearpage
%4

\clearpage 
%5

\clearpage
%6

\clearpage
%7
\clearpage

%8

\clearpage
%9
\clearpage
%10
\clearpage
%11
\clearpage
%12
\clearpage
%13
\clearpage
%14

\clearpage
%15
 
 \clearpage
%16 
 \begin{figure}[h]
\centering
\vspace*{-4.5em}
\includegraphics[width=0.85\linewidth]{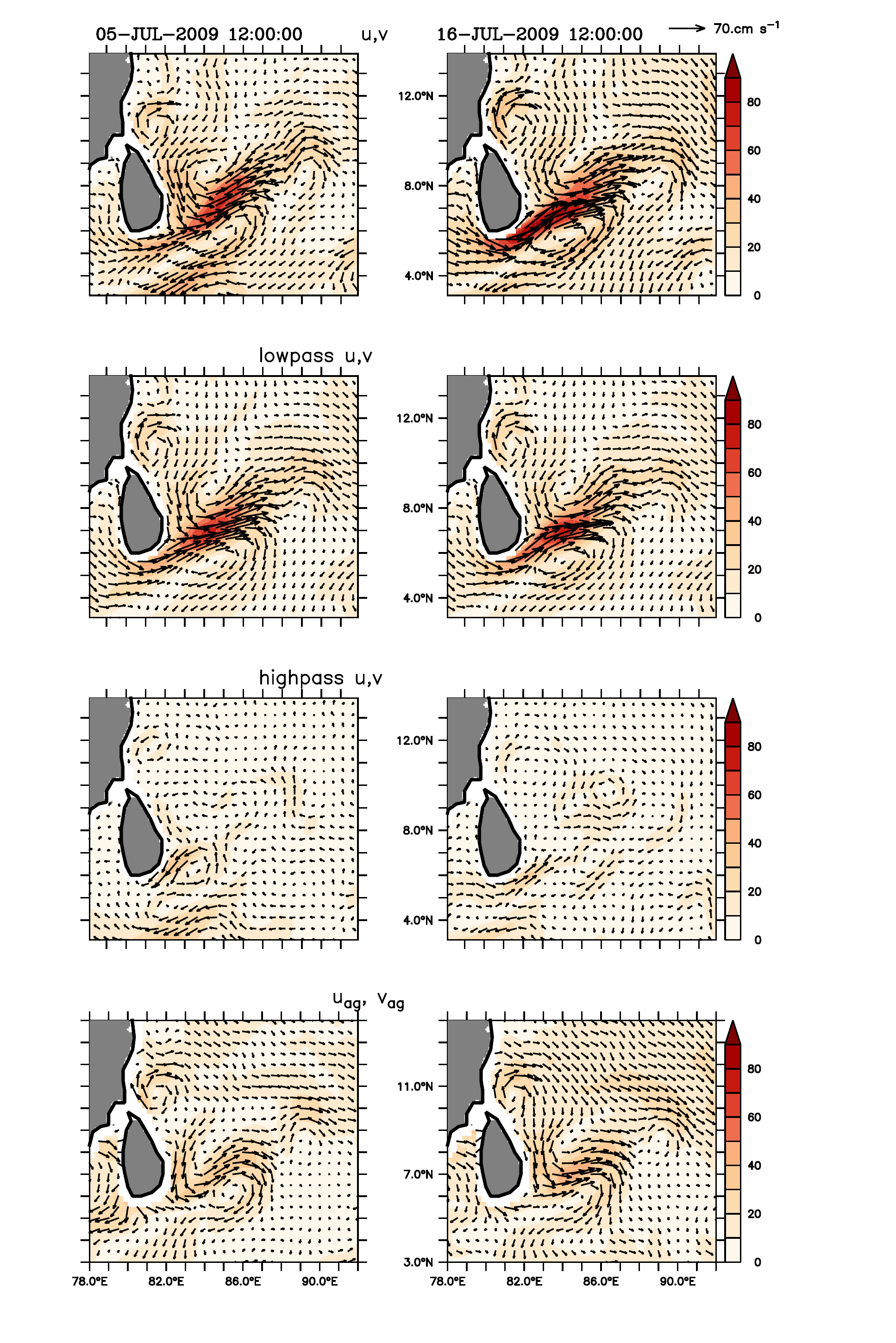}
\caption{Snapshots of current vectors overlaid over speed (shading, cm/s) averaged over top 30m on 5/07/2009 (left panel) and 16/07/2009 (right panel). First row represents the model currents, second row represents the 120 day low-pass current components, third row shows the 120 day high-pass current components and the fourth row shows the ageostrophic current components. Vectors are of same length in all panels and the scale vector is shown in top panel.}
\label{fig:ageos}
\end{figure}

\clearpage
 %17

\clearpage
%18

\end{document}